\pdfoutput=1
\documentclass{aastex62_arx}

\usepackage{amsmath,graphicx}
\usepackage{amsthm,amssymb}
\usepackage{verbatim}
\usepackage{epstopdf}
\usepackage{tikz} 
\usetikzlibrary{shapes,arrows}
\usepackage[toc,page]{appendix}
\usepackage{subcaption}
\usepackage{rotating}

\newcommand{\EE}{\mathbb{E}}
\newcommand{\var}{\text{Var}}

\tikzstyle{block} = [draw, fill=white, rectangle, 
    minimum height=3em, minimum width=7em]
\tikzstyle{sum} = [draw, fill=white, circle, node distance=1cm]
\tikzstyle{input} = [coordinate]
\tikzstyle{output} = [coordinate]
\tikzstyle{pinstyle} = [pin edge={to-,thin,black}]

\begin{document}

\title{UNCERTAINTY QUANTIFICATION IN SUNSPOT COUNTS}

\author{Sophie Mathieu}
\author{Rainer von Sachs}
\author{Christian Ritter}
\affiliation{ Universit\'{e} catholique de Louvain\\
	ISBA\\
	Louvain-la-Neuve, Belgium\\}
	\nocollaboration

\author{V\'{e}ronique Delouille}
\author{Laure Lef\`{e}vre}
\affiliation{Royal Observatory of Belgium\\
	Solar physics and Space Weather department\\
	Brussels, Belgium}
\nocollaboration




\begin{abstract}
Observing and counting sunspots constitutes one of the longest-running scientific experiment, with first observations dating back to Galileo and the invention of the telescope around 1610. Today the sunspot number (SN) time series acts as a benchmark of solar activity in a large range of physical models.  An appropriate statistical modelling, adapted to the time series' complex nature, is however still lacking.

In this work,  we provide the first comprehensive uncertainty quantification analysis of sunspot counts. Our interest lies in the following three components: the number of spots ($N_s$), the number of sunspot groups ($N_g$), and the composite $N_c$, defined as $N_c:=N_s+10N_g$. Those are reported by a network of observatories around the world, and are corrupted by errors of various types. We use a multiplicative framework to provide, for each of the three components, an estimation of their error distribution in various regimes (short-term, long-term, minima of solar activity). We also propose a robust estimator for the underlying solar signal and fit a density distribution that takes into account intrinsic characteristics such as over-dispersion, excess of zeros, and multiple modes. The estimation of the solar signal underlying the composite $N_c$ may be seen as a robust version of the International Sunspot Number (ISN), a quantity widely used as a proxy of solar activity. Therefore our results on $N_c$ may serve to characterize the uncertainty on ISN as well. 

Our results paves the way for a future monitoring of the observatories in quasi-real time, with the aim to alert the observers when they start deviating from the network and prevent large drifts from occurring in the network.

\end{abstract}


\section{INTRODUCTION}
\label{sec:introduction}
\subsection{The International Sunspot Number (ISN)}

On white light images, sunspots are visible as dark areas. They correspond to regions of locally enhanced magnetic field and act as an indicator of changing solar activity over time. They have been observed and counted since the invention of the telescope at the beginning of the 17th century. As such, the counting of sunspots constitutes one of the ``longest-running scientific experiment''~\citep{Owens2013}. In 1848, J.R. Wolf from Z\"{u}rich Observatory created an index, denoted $N_c$, of solar activity by summing up the total number of sunspots $N_s$ with ten times the total number of sunspots groups $N_g$ on a daily basis:
 \begin{equation}
 \label{E:Nc}
N_c = 10 N_g + N_s
\end{equation}

\begin{figure}[!hb]
	\centering
		\includegraphics[scale=0.8]{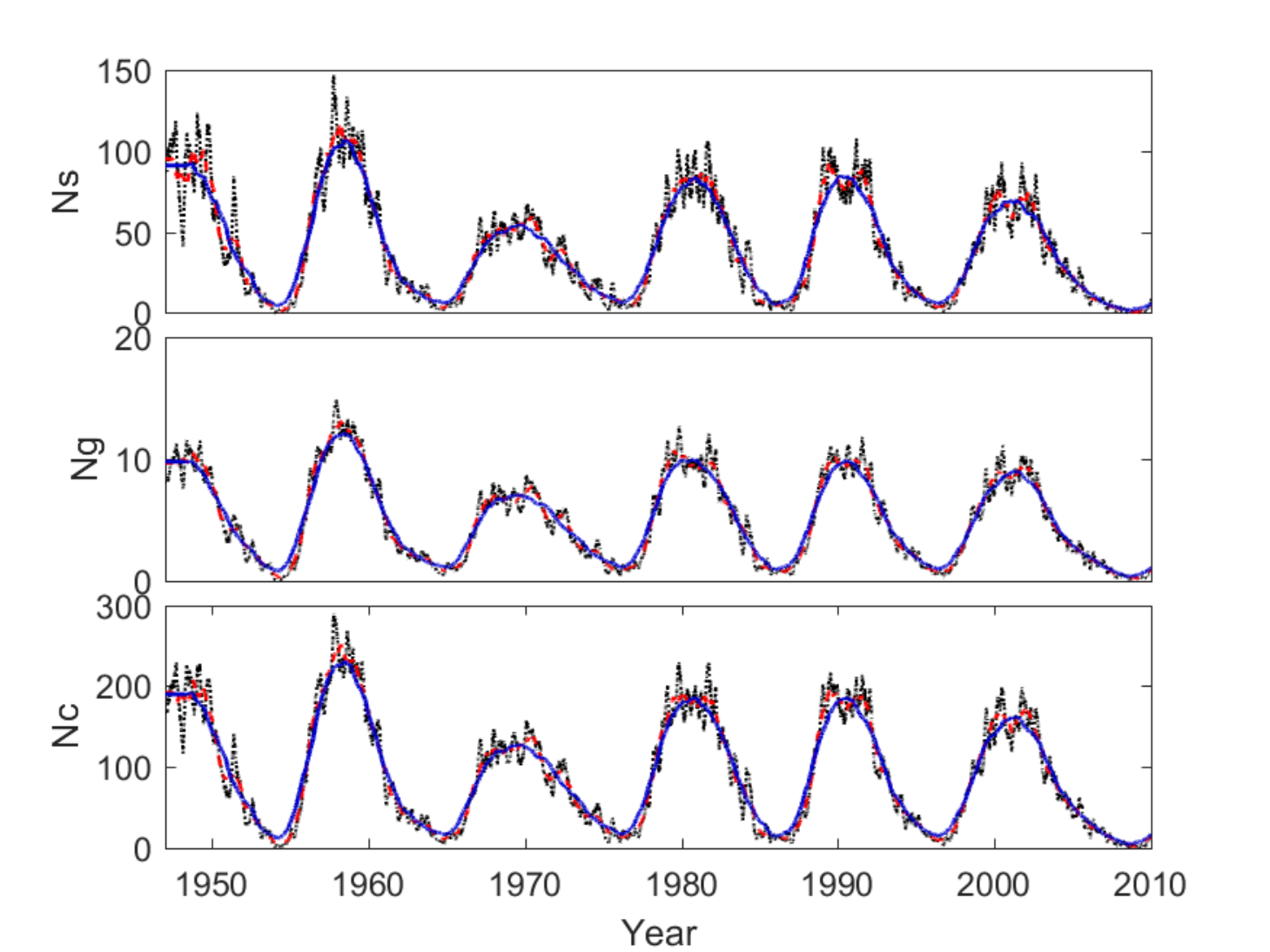}
	\caption{\small{Time evolution during 1947-2013 of the median values across 21 observing stations (cf. Table \ref{tab:station}) for the sunspot counts: (top) $N_s$, (center) $N_g$,  and  (bottom) $N_c$.  The data are averaged over 81 days (black dotted line), 1 year (red dashed line) and 2.5 years (blue plain line).}}
\label{fig:ss}
\end{figure}

Figure~\ref{fig:ss} displays smoothed averages of the median value of these three quantities across a set of 21 observatories (also called \lq stations') chosen for the present study (see Section~\ref{sec:data}). Modelling the statistics of $N_c$ is far from trivial, as this quantity jumps from zero to eleven when a sunspot appears on the Sun ($N_s =1, N_g=1,$ and thus $ N_c=11$). By construction, each active region appears thus twice in $N_c$. The multiplication factor in~Equation(\ref{E:Nc}) was introduced by J.R. Wolf to put the number of groups on the same scale as the number of spots. Indeed,  during this historical period, a group contained on average ten spots~\citep{Izenman1985}. Note that in recent solar cycles, the average number of spots per group is rather around six.  

The index $N_c$, or rather the formula behind it, is at the basis of the International Sunspot Number (ISN). The ISN is distributed through the World Data Center Sunspots Index and Long-term Solar Observations (WDC-SILSO) \footnote{\url{http://www.sidc.be/silso/}}. The $N_c$s from each observing station in the SILSO network are collected and rescaled, i.e. multiplied by a factor $k$, to compensate for their differing observational qualities. $N_c$s are then combined on a monthly basis to produce the ISN~\citep{Clette2007}, which constitutes  the international reference for modelling solar activity on the long-term. Despite the fact that it is arguably the most intensely used times series in all of astrophysics~\citep{Hathaway2010}, its historical part suffers from a number of errors and inconsistencies~\citep{all_corrections}. Even the most recent part (1981-now) lacks proper error modelling and uncertainty quantification, see Section~\ref{ssec:origin} below.

\subsection{ISN: Origin and computation}
\label{ssec:origin}

\begin{figure}[htbp]
	\centering
		\includegraphics[width=12cm, height=6.5cm]{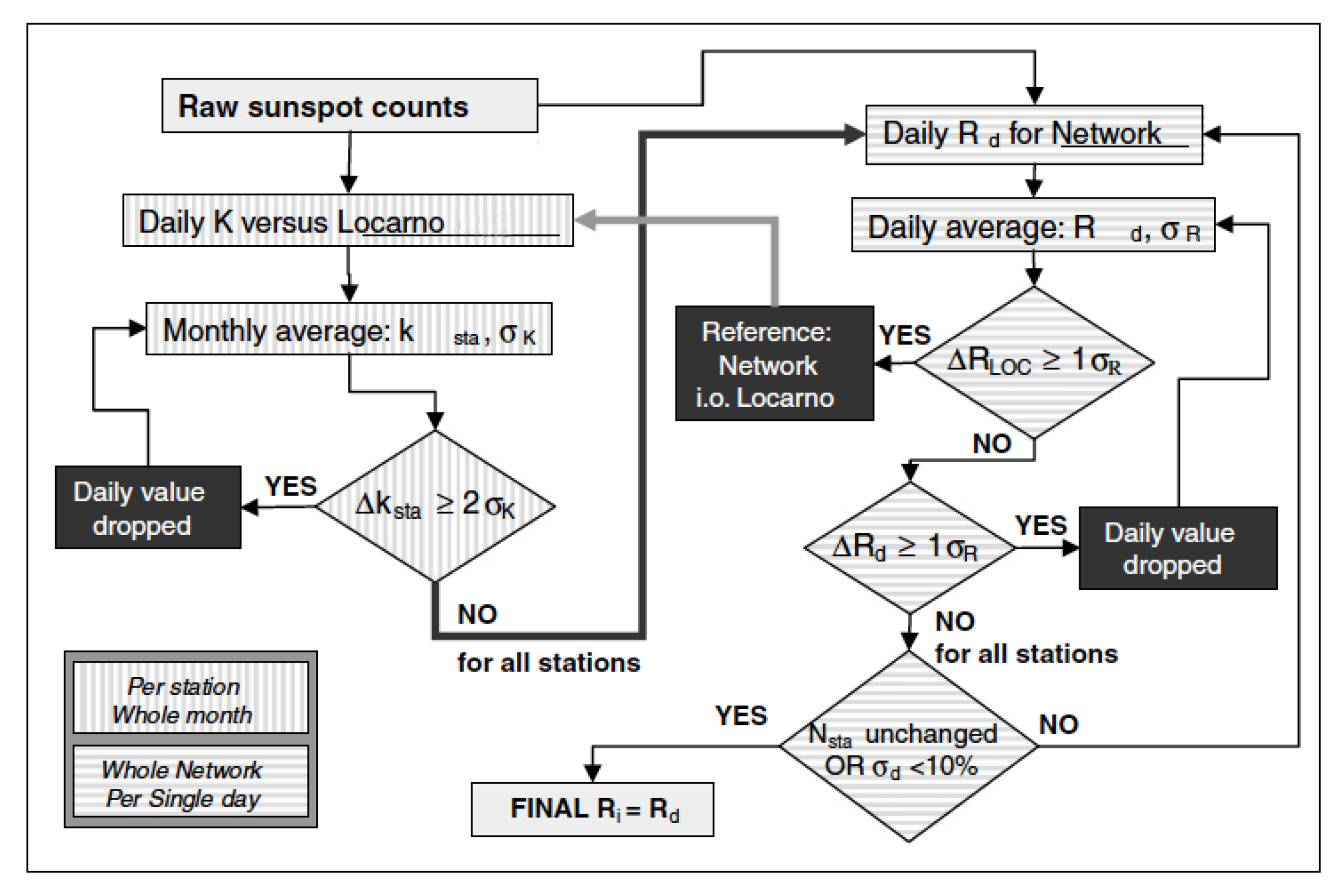}
	\caption{Flowchart of the WDC-SILSO data import procedure, illustrating the succession of hierarchical tests applied to raw observing reports (adapted from~\cite{Clette2007}).}
\label{fig:Kfact}
\end{figure}

In order to place our work in context, we first describe how the ISN is currently obtained  at the WDC-SILSO center. $N_s$ and $N_g$ are entered through an interface (\url{www.sidc.be/WOLF}) and stored in a database. The main processing is described in~\cite{Clette2007} and we summarize a more recent version of it in Figure~\ref{fig:Kfact}. The processing uses a pilot station, here the Locarno station, as a reference. It compares the values obtained by a station $i$ to the pilot station via a scaling factor $k_i$, often referred as the \lq $k$-coefficient': 
\begin{equation}
 k_i(t)=\frac{pilot(t)}{Y_i(t)},
\label{E:k_factor}
\end{equation}
where $Y_i(t)$ is the composite index of station $i$, observed at time $t$ (expressed in days) and $pilot(t)$ is the value of the pilot station.
The monthly scaling factors are computed from a sigma-clipping mean of Equation~(\ref{E:k_factor}), i.e. values differing by more than two standard deviations from the mean are eliminated from the computation process.
This processing still suffers from its historical heritage, summarized in Table 1 of~\cite{Dudok2016}. For example, this table shows that between 1926-1981 (when the Sunspot collection center was in Zurich) there were several standard observers, and no pilot station. 
As an index derived from count data, $N_c$ (or $N_s$ and $N_g$) does not necessarily follow a Gaussian distribution~\citep{Dudok2016, Usoskin2003, Vigouroux1994}. A processing based on  sigma-clipping  is  thus not fully adapted, but still undoubtedly better than what was done during the Zurich era.  Finally, some steps in the processing date back from the mid-19th century (when J.R. Wolf introduced the sunspot index) and have not been upgraded when the collection and preservation centre was moved from Zurich to Brussels in 1981. There were two reasons for this: (1) the new curators of the International Sunspot Number wanted to keep the uniformity of the series, and (2) the numerical tools available at that time were limited.

The WDC-SILSO team is currently working on improving the ISN computation and coordinates an important community effort to correct past errors. Such effort includes amongst others: work by   a team from the International Space Science Institute (ISSI) on recalibration of the SN\footnote{\url{http://www.issibern.ch/teams/sunspotnoser/}}, organization of sunspot workshops\footnote{\url{https://ssnworkshop.fandom.com/wiki/Home}}, editorial work for a Solar Physics topical issue on SN recalibration \footnote{\url{https://link.springer.com/article/10.1007/s11207-016-1017-8}}.

\subsection{Previous works on SN uncertainty quantification}
\label{ssec:previous}
Long-term analyses started with models of the shape of the sunspot number time series~\citep{Stewart1940, Stewart1938}. They pursued the works by M. Waldmeier himself~\citep{Waldmeier1939} who tried to understand the solar cycle and predict upcoming cycles. Later on,~\cite{Morfill1991} investigated the short-term dynamical properties of the SN series using a Poisson noise distribution superimposed on a mean cycle variation.  \cite{Vigouroux1994} also uses a Poisson distribution to approximate the dispersion of daily values of the SN  at different regimes of solar activity. \cite{Usoskin2003} develops a reconstruction method for sparse daily values of the SN and models the monthly number of groups corresponding to a certain level of daily values by a Poisson distribution. 
\cite{Schaefer1997} emphasizes the need for error bars on the AAVSO sunspot series\footnote{\url{https://www.aavso.org/category/tags/american-relative-sunspot-numbers}} and more recent results in~\cite{Dudok2016} present a first uncertainty analysis of the short-term error, through time domain errors and dispersion errors among observing stations, still assuming a Poisson distribution. In~\cite{Dudok2016} however, the authors uncover the presence of over-dispersion in the SN, and approximate the SN by a mix of a Poisson and a Gaussian distribution in an additive framework. Although non-Poissonian, this additive model fails to capture some of the characteristics of sunspot data. 
\cite{Oh2012} on the other hand uses a multiplicative model to simulate sunspot counts in view of assessing the dependency of correction factors on the solar cycle.

\subsection{Motivation and contribution}
\label{ssec:isn-model}
Our goal in this work is to go beyond the above-mentioned historical heritage by developing a comprehensive uncertainty quantification model for the count data $N_s$,  $N_g$ and $N_c$. 
These quantities are subject to different types of errors and do not behave exactly like Poisson random variables: (1) they experience more dispersion than  the Poisson distribution, (2) they are not independent from one day to another (since sunspots can last from several minutes to several months on the sun) and (3)  they exhibit a large number of zeros due to periods of minimal solar activity.

Our contribution is two-fold. First, we develop  robust estimators for the physical solar signal, denoted \lq true' signal in this paper, underlying $N_s$, $N_g$ and $N_c$.  We propose a model for their densities that takes into account characteristics such as overdispersion and large number of zero counts. Our processing and estimators are robust to missing values and do not require to fill-in missing observations, contrarily to previous studies.

Second, we propose an uncertainty model that is motivated by first studies in~\cite{Dudok2016} and that works within a \emph{multiplicative} framework. Our model distinguishes three error types. The {\it short-term error} accounts for counting errors and  variable seeing conditions from one station to another (e.g.~weather, atmospheric turbulence), whereas the  {\it long-term error} provides an overall bias in the number of spots (e.g.~gradual ageing of the instrument or observer). Finally, a third error type aims at modeling inaccuracies occurring at solar minima, and helps differentiating true from false zero counts.
As an illustration, the short-term variations coming from the solar variability and the observational errors are clearly visible in Figure~\ref{fig:ss}, superimposed on the approximate seasonality of the eleven-year solar cycle. 

In future prospects, our work paves the way for a more robust definition of the ISN.
Indeed, the analysis of the different error types allows studying the stability of observatories involved in the computation of ISN. Our study lays the ground for a future monitoring of all active stations within the SILSO network in quasi-real time, with the aim to: (1) define a stable reference of the network, (2) alert the observers when they start deviating from the network and (3) prevent large drifts from occuring. A new ISN could then be defined from a stable reference rather than a single pilot station, and benefit from the robust estimators and procedures (including rescaling of observing stations, see Section~\ref{sec:pre-processing}) developed  in this work. 

Our paper is structured as follows. Section~\ref{sec:data} introduces the dataset considered. The uncertainty model is presented in Section~\ref{sec:model} while Section~\ref{sec:pre-processing} details the pre-processing of the data. 
Section~\ref{sec:solar-signal-estimation} provides the estimators (or proxy) for the \lq true' solar signal underlying $N_s$, $N_g$ and $N_c$  as well as their densities. 
Finally, Section~\ref{sec:errors} displays our results on quantification of the different error types, as well as a 
first stability analysis that takes into account both short and long-term variability. 

\section{DATA}
\label{sec:data}

Similarly to what is done in~\cite{Dudok2016}, we study a subset of 21 stations, whose main characteristics are listed in Table~\ref{tab:station}. The period under study goes from 1947 January 1 till 2013 December 31. It ranges from the maximum of solar cycle (SC) 18 until the ascending phase of SC 24\footnote{\url{https://en.wikipedia.org/wiki/List_of_solar_cycles}} and covers thus almost six solar cycles.

Table \ref{tab:station} summarizes properties of the stations such as their location, name, type (amateur vs professional, individual vs team),  observing period, percentage of missing values, and  mean scaling factor (level) with respect to the network over the period studied.
The procedure that was used to  compute the mean scaling-factors will be described in Section~\ref{sec:pre-processing}. This mean scaling-factors may be viewed as an indication of the general level of counts recorded by the station  as compared to the median of the network.  Thus $\mbox{\emph{level}} = 1$ corresponds to a station that observes the same number of spots than the median of the network. For example, Locarno (LO) with a level of $1.26$ observes in general around 20 \% more spots than the others.  

The location of the observatories gives an indication of the weather conditions of the stations and might explain part of the missing values. 
Moreover, the type of observatory usually impacts the quality of observations and the length of observing periods: 
an individual might experience less short-term variability than a team (alternating the observer from one week to another) and/or amateurs may have shorter observing periods than professionals. 

\begin{table}[ht]
\caption{Main characteristics of the subset of stations}
\begin{center}
\begin{tabular}{|c|c|c|c|c|c|c|c|c|}
\cline{1-9}
ID & Full & Location & Prof. vs & Individual & Observing & Level & \%  & \% Obs.  \\
 & name &  & amateur & or team  & period  &  & Obs. & period \\
    \hline
 A3 & Athens Observatory & Athens (Greece) & Prof. & team & 1949-1995 & 1.039 & 30.16 & 44.01\\
 BN-S & WFS Observatory (SONNE) & Berlin (Germany) & Am. & team & 1965-2013 & 1.179 & 23.50 & 32.74 \\
 CA & Catania Observatory & Catania (Italy) & Prof. & team &1950-2019 & 1.039 & 61,87 & 64.80\\
 CRA & T. Cragg (AAVSO) & Australia & Am & indiv. & 1947-2009 & 0.904 & 72,43& 77.44\\
 FU & Fujimori & Nagano (Japan) & Am & indiv. &1968-2019 & 1.055 & 45,73& 67.32\\
 HD-S & Hedewig (SONNE) & Germany & Am & indiv. &  1967-2013 & 0.931 & 25,42& 36.96\\
 HU & Public Observatory & Hurbanovo (Slovakia) & Am & team & 1969-2019& 1.004 & 35,452& 52.80\\
 KH & KOERI & Kandilli (Turkey) & Prof. & team & 1967-2019 & 0.968 & 48,81& 51.38 \\
 KOm & I. Koyama & Japan & Am & indiv. & 1947-1996 & 1.052 & 40,18& 54.84\\
 KS2  & Observatory & Kislovodsk (Russia) & Prof. & indiv. & 1954-2019& 1.057 & 85,96 & 95.98\\
 KZm & University of Graz & Kanzelhohe (Austria) & Prof. & team &  1944-2019& 1.110 & 74,23 & 74.24\\
 LFm & H. Luft &  New York (USA) & Am & indiv. & 1944-1988& 0.985 & 34,06  & 54.68\\
 LO & Specola Solare & Locarno (Switzerland) & Prof. & indiv. & 1958-2019 & 1.260 & 68,27 & 81.68 \\
 MA & & Manila (Philippines)  & Prof. & team & 1971-1988 &  1.023 & 20,85 & 78.69\\
 MO  & Mochizuki (Urawa) & Saitama (Japan) &  Am & indiv. & 1978-2019&  1.073 & 35,51 & 66.09\\
 PO & Observatory & Postdam (Germany) & Prof. & team & 1955-1999 &  0.991 & 22,12 & 29.73\\
 QU & PAGASA weather Bureau & Quezon (Philippines)  & Prof. & indiv. & 1957-2019 &  0.829 & 45,46 & 53.83\\
 SC-S  & Schulze (SONNE) & Germany & Am & indiv. & 1960-2007 &  0.943 & 23,32 & 33.16\\
 SK & Skalnate Pleso Obs. & Vysoke Tatry (Slovakia) & Prof. & team &1950-2012 & 0.992 & 37,95 & 40.75\\
 SM & & San Miguel (Argentina) & Prof. & team &1967-2013 & 1.220 & 39,09 & 56.34\\
 UC & USET & Uccle (Belgium) & Prof. & team & 1949-2019 & 0.991 & 57,00 & 59.64 \\
 \hline 
\end{tabular}
\end{center}
\caption*{\small{Main characteristics on the set of 21 stations used in this study:  acronym (ID), name (Full Name), location (Location), type of observatory (professional vs amateur), type of observer (individual or team), observing period (Observing period), averaged scaling factor with respect to the network over the studied period (Level), percentage of observations on the full period studied (\% Obs.) and on their observing period (\% Obs. period). Note that SONNE and AAVSO (mentioned in column \lq Full Name') are two distinct networks of observing stations which are not members of the WDC-SILSO network and hence are not used to produce the ISN.}}
\label{tab:station}
\end{table}
 
Our dataset contains the daily number of spots $N_s$, groups $N_g$ and the composite $N_c$ observed in each of the 21 stations.  As Table~\ref{tab:station} indicates, the data present an important amount of missing values due to: weather conditions preventing stations from observing, periods of instrument maintenance or definitive closures or births of new observatories. 

\section{MODEL}
\label{sec:model}

In this section, we present step by step our uncertainty model. 
It characterizes the observations of the stations (either for the number of spots $N_s$, groups $N_g$ or composite $N_c$) in a multiplicative framework and involves different types of observing errors  as well as a quantity generically denoted by $s(t)$, for $N_s$, $N_g$ and $N_c$. $s(t)$ is a latent variable representing the \lq true' solar signal, i.e. the actual number of spots $N_s$, groups $N_g$ or composite $N_c$ lying on the Sun. It cannot be directly observed as the counts of the stations are corrupted by different error sources.
Our goal is to estimate the distribution of the \lq true' solar signal and of the errors degrading it. In particular, we are interested in the mean and the variance of these distributions but also in higher order moments since the estimated densities are far from Gaussian.

 The mean of $s(t)$, denoted by $\mu_s(t)$, will be estimated in Section~\ref{sec:solar-signal-estimation} based on the entire network (to be robust against errors of an individual station) and will be used as a proxy for $s(t)$ in the remaining part of the article.
 Since our model is multiplicative, a good estimation of $\mu_s(t)$ is the key to get access to the multiplicative errors, cf. Equation~(\ref{E:model}). Moreover, a precise estimation of the mean level of an individual station is required for future monitoring, and this depends on the accuracy of the estimation of $s(t)$.
 
 We use a model that is conditional on the latent $s(t)$ and decomposes the observations along two regimes: when $s(t)=0$ (solar minima) and when $s(t)>0$  (outside periods of minima), see Section~\ref{ssec:conditional}. This allows introducing, outside of minima, a model with short-term observing errors and long-term drifts, cf. Section~\ref{ssec:short-long}. A specific error model is then developed for  periods of solar minima in Section~\ref{ssec:minima}, and the complete model is shown in~\ref{ssec:general}. Finally,  Section~\ref{ssec:excess} introduces the Hurdle model in order to fit distributions exhibiting an excess of zero values, as it is the case here due to the presence of solar minima and observing errors. 

\subsection{Conditional model}
\label{ssec:conditional}
The observed counts are studied in two distinct situations: when there are sunspots ($s>0$) and when there are none ($s=0$). This separation is motivated by the idea that the absence or the presence of sunspots are lead by a series of phenomena involving complex dynamo processes in the solar interior, and which can be modelled by a latent variable with two states. 
This analysis will lead to a better understanding of the observations and allows differentiating the \lq true' zeros of the counting process from the \lq false' zeros that occur when a station reports zero sunspot count in presence of one or more spots on the Sun. 

Let $Y_i(t)$ represent either the number of spots, groups or composite actually observed (raw, unprocessed data).  The index $1 \leq i\leq N$ denotes the station, and $1 \leq t \leq T$ is the time. 
The probability density function (PDF) of $Y:=Y_i(t)$ may be decomposed as:

\begin{equation}
\label{E:conditional}
\begin{split}
\mathbf{P}(Y=0)= \quad & \overbrace{P(Y=0|s(t)> 0) P(s(t) > 0)}^{1}  \\
 & + \underbrace{P(Y=0 |s(t)=0) P(s(t)=0)}_{2}   \\
\mathbf{P}(Y\geq y)= \quad & \overbrace{P(Y \geq y | s(t)>0)P(s(t)>0)}^{3} \\
& + \underbrace{P(Y\geq y |s(t)=0) P(s(t)=0)}_{4} \ \text{for} \  y>0. \\
\end{split}
\end{equation}
Terms \lq 1' and \lq 3' in Equation~(\ref{E:conditional}) represent the short-term error in presence of one or more sunspots. Term \lq 1' reflects a situation where no sunspots are reported while there are actually some spots on the Sun (\lq false' zeros or observational errors due e.g. to a bad seeing), and leads to an excess of zeros in short-term error distribution.

Term \lq 2' captures the  \lq true' zeros (no sunspot and no sunspot reported) while term \lq 4' reflects a situation where the station reports some sunspots when there are no sunspot on the Sun.  Term \lq 4' is neglected outside of solar minima periods.
Together, these two terms form the distribution of the error at minima, which has an excess of \lq true' zeros and a tail modelling the errors of the stations and the short-duration sunspots.

\subsection{Short-term and long-term errors}
\label{ssec:short-long}

Results in~\cite{Dudok2016} evidence a short-term, rapidly evolving, dispersion error across the stations that accounts for counting errors and variable seeing conditions. We define a similar term allowing a possible station-dependence and we denote it $\epsilon_1(i,t)$. Assuming $\EE(\epsilon_1(i,t))=0$, where $\EE$ is the expectation sign, our interest lies in modelling its variance and its tail to study the short-term variability of the stations. \\
Next, we introduce $\epsilon_2(i,t)$ to handle station-specific long-term errors such as systematic biases in the sunspot counting process. We want to estimate its mean, $\mu_2(i,t)$, and detect if this mean experiences sudden \emph{jumps} or \emph{drifts} on longer time-scales.

Both $\epsilon_1(i,t)$ and $\epsilon_2(i,t)$ are multiplicative errors, as an observer typically makes larger errors when $s(t)$ is higher~\citep{Oh2012}. Assembling these two types of errors, we propose the following noise model outside of solar minima:
\begin{equation}
\label{E:model-nz}
\begin{split}
 Y_i(t)=(\epsilon_1(i,t)+\epsilon_2(i,t)) s(t) & \  \text{when} \  s(t)>0.
\end{split}
 \end{equation}

\subsection{Errors at solar minima}
\label{ssec:minima}

Let $\epsilon_3$ denote the error occurring during minima of solar activity, when there exists extended periods with no or few sunspots. We assume the error $\epsilon_3$ to be significant when there are no sunspots ($s(t)=0$) and otherwise negligible in order to not interfere with the errors $\epsilon_1$ and $\epsilon_2$. $\epsilon_3$ captures effects like short-duration sunspots and non-simultaneity of observations between the stations. 
At solar minima, the model becomes: 
\begin{equation}
\begin{split}
 Y_i(t)= \epsilon_3(i,t) & \  \text{when} \  s(t)=0.
\end{split}
 \end{equation}

\subsection{General model}
 \label{ssec:general}
Combining the three error types, we may write our uncertainty model in a compact and generic way as follows: 
\begin{equation} 
\begin{split}
Y_i(t) = \left\{ \begin{array}{ll} (\epsilon_1(i,t)+\epsilon_2(i,t))s(t) & \text{if} \  s(t)>0 \\ 
\epsilon_3(i,t) & \text{if} \  s(t)=0. \\ \end{array} \right.  
\end{split}
 \label{E:model}
 \end{equation}

We assume the random variables (r.v.) $\epsilon_1$, $\epsilon_2$ and $\epsilon_3$ to be continuous, and the r.v.  $s$, $\epsilon_1$, $\epsilon_2$ and $\epsilon_3$ to be jointly independent.  Although the \lq true' number of counts $s(t)$ is discontinuous, its product with a continuous r.v. ($\epsilon_1+\epsilon_2$) remains continuous.
This is consistent with the fact that, after preprocessing, the observed data  $Y_i(t)$ may be modelled by a continuous r.v.

\subsection{Excess of zeros}
\label{ssec:excess}
All terms in Equation~(\ref{E:model}) exhibit an excess of zeros, that is, an unusual local peak in the density at zero, due to solar minima periods. As the solar minimum is an important part of a solar cycle, the zeros must be properly treated. Specific models such as the zero-altered (ZA) or the zero-inflated (ZI) two part distributions may be used for this purpose~\citep{count_ecology,count_data}. The main difference between both models is that the ZI distribution allows the zeros to be generated by two different mechanisms contrarily to the ZA model which treats all zeros in the same way. 

As \lq true' and \lq false' zeros do not appear together in a single term of Equation~(\ref{E:model}), we find appropriate to work with the ZA two-part model (also called \lq Hurdle' model) and denote its density by $f(x)$. In this model, the zero values are modelled by a Bernoulli distribution $f_0(x)=b^{1-x}(1-b)^{x}$ of parameter $b$. Non-zero values follow a distribution described generically by $f_1(x)$, either another discrete distribution (in case of modelling the counts $\mu_s(t)$ in Section~\ref{sec:solar-signal-estimation}) or a continuous distribution for $\epsilon_1$ and $\epsilon_3$ in Section~\ref{sec:errors}: 
\begin{equation}
\label{E:hurdle}
 f(x) = \begin{cases}
             f_{0} (0)  = b  & \text{if } x = 0 \\
             (1-f_{0}(0))\frac{f_{1}(x)}{1-f_{1}(0)} = (1-b) \  \frac{f_{1}(x)}{1-f_{1}(0)} & \text{if } x > 0.
          \end{cases} 
\end{equation}
The ZA distribution will be used to model the estimated densities of $\mu_s(t)$ in Section~\ref{sec:solar-signal-estimation}, and of $\epsilon_1(i,t)$ and $\epsilon_3(i,t)$ in Section~\ref{sec:errors}.

\section{PRE-PROCESSING}
\label{sec:pre-processing}

 Due to the different characteristics of the observing means (telescope aperture, location, personal experience etc.), each station has its own scaling. These differences mainly impact the count of small spots, which cannot be observed with low-resolution telescopes, and the splitting of complex groups, where the personal experience of the observer matters.  A pre-processing is thus needed to rescale all stations to the same level when comparing stations on the short-term and at solar minima. It is also required to compute a robust estimator of the solar signal based on the entire network. For the analysis of long-term errors however, the pre-processing will not be applied, as it would suppress long-term drifts that we want to detect. Our proposed pre-processing is robust to missing values and proceeds in two steps. 

First, we compute the \lq time-scale', that is,  the duration of the period where the scaling-factors are assumed to be constant. It is a trade-off between short periods and long periods: the former tends to standardize the observations of the stations, thereby suppressing any differences between the observers,  whereas the latter may be too coarse to correct for important changes in observers and instruments. A statistical-driven study based on the Kruskal-Wallis test~\citep{Kruskal_Wallis} shows that the appropriate time-scale varies with the stations and with the type of counts $N_s$, $N_g$, and $N_c$, see Appendix~\ref{sec:AA} for a full description of the test. This time-scale may also evolve over time when a station is constant over several months before suddenly deviating from the network.   
However, to avoid introducing potential bias between the stations, we use the \emph{same} time-scale, generically denoted by $\tau^\star$, for all stations over the entire period studied. The selected values of $\tau^\star$ are: 8 months for $N_s$, 14 months for $N_g$ and 10 months for $N_c$.  We note that that these periods are close to the twelve-months period chosen by J.R. Wolf to compute the historical version of the scaling-factors. 

Second, having determined the time-scale $\tau^\star$, we compute the scaling-factors  using ordinary least-squares regression (OLS) as follows.
Recall that $Y_i(t)$ represents either the number of spots, groups or composite actually observed in a station $i$, $1 \leq i\leq N$ at time $t$, $1 \leq t \leq T$ (daily values). For convenience, we re-arrange the time by an array of two indices $t=(t_1,t_2)$, where $1 \leq t_1 \leq \tau^\star$ and $1 \leq  t_2 \leq T/\tau^\star$. Thus, $t_1$ corresponds to the index of an observation inside a block of length $\tau^\star$ and $t_2$ is the index of the block. 

Let $\vec{Y}_{i,t_2}:=[Y_i((t_1,t_2))]_{1 \leq t_1 \leq \tau^\star}$ denote the vector of the daily observations in station $i$ on block $t_2$ of length $\tau^\star$ and 
$\vec{X}_{i,t_2}:=[\underset{1 \leq i\leq N}{\text{med}} Y_i((t_1,t_2))]_{1 \leq t_1 \leq \tau^\star}$ be the vector containing the daily values of the median of the network, also of length $\tau^\star$. The scaling-factors are computed using the slope of the $OLS(\vec{Y}_{i,t_2}|\vec{X}_{i,t_2})$ regression: 
 \begin{equation}
 \kappa_i(t_2)=(\vec{X}_{i,t_2}^T \vec{X}_{i,t_2})^{-1} \vec{X}_{i,t_2}^T \vec{Y}_{i,t_2}.
\label{E:ki}
\end{equation}

The new definition of the scaling-factors in Equation~(\ref{E:ki}) is a robust version of a ratio between the observations of the stations and the median of the network. It is similar to the definition of the historical $k$ in Equation~(\ref{E:k_factor}), where the median of the network replaces the single pilot-station as the reference level.
The rescaled data, denoted $Z_i$ in the sequel, are defined as:
\[
Z_i(t) = \frac{Y_i(t)}{\kappa_i(t)},
\]
where the reference appears now in the denominator. In a sense, the ratio in Equation~(\ref{E:k_factor}) is inverted in order to limit the problem of dividing by zero whenever the stations observe no spots.  We explored other methods such as orthogonal regression (also called Total Least Squares) and $OLS(\vec{X}_{i,t_2}|\vec{Y}_{i,t_2})$~\citep{TLS_OLS}.
We choose the $OLS(\vec{Y}_{i,t_2}|\vec{X}_{i,t_2})$ method since it leads to the smallest Euclidean and Total variation distances between the median of the network and the individual  stations.

\section{SOLAR SIGNAL ESTIMATION}
\label{sec:solar-signal-estimation}

\subsection{Choice of the estimator}
\label{ssec:filtering}

\begin{figure}[h]
\begin{tikzpicture}[auto, node distance=2cm,>=latex']

    \node (input) {$M_t$};
    \node (asc) [block,right of=input,node distance=2cm] {$A(M_t)$};
     \node (fft) [block,right of=asc,node distance=3cm] {FFT(Q)};
     \node (ths) [block,right of=fft,node distance=3cm] {Wiener filter};
     \node (ifft) [block,right of=ths,node distance=3cm] {IFFT($\phi'$)};
     \node (iasc) [block,right of=ifft,node distance=3cm] {$A^{-1}(Q')$};
      \node (output) [right of=iasc, node distance=2cm] {$\hat{\mu}_s(t)$};
      
    \draw [->] (input) -- node[name=u1] {} (asc);
     \draw [->] (asc) -- node[name=u2] {Q} (fft);
      \draw [->] (fft) -- node[name=u3] {$\phi$} (ths);
       \draw [->] (ths) -- node[name=u4] {$\phi'$} (ifft);
       \draw [->] (ifft) -- node[name=u5] {Q'} (iasc);
        \draw [->] (iasc) -- node[name=u6] {} (output);
       
\end{tikzpicture}
\caption{\small{ Block diagram of the $T$ procedure defining the solar signal estimator $\hat{\mu}_s(t)$, where $M_t=\underset{1 \leq i\leq N}{\text{med}} Z_i(t)$ is the median of the network. An Anscombe transform is first applied on the median, and missing values are imputed. 
Then, a fast Fourier transform (FFT) is used to convert the signal to a power spectrum in the frequency domain, followed by an attenuation from a Wiener filter.
A step function cancels the amplitude of the frequencies corresponding to the periods inferior to seven days (low-pass filter). The threshold at seven days is selected from \cite{Dudok2016}, Fig 5. It is the smallest visible time-scale of the signal, corresponding to the weekly shift of some observatories.
Finally, an inverse FFT and an inverse Anscombe transform are applied to the signal.
}}
 \label{Fig:filtering}
\end{figure}
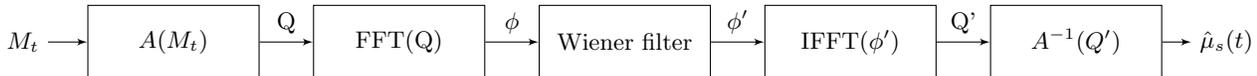

To use Equation~(\ref{E:model}), we need an estimate of a proxy for $s(t)$. We choose this proxy to be the mean of $s(t)$, denoted $\mu_s(t)$.
We propose as a robust estimator for $\mu_s(t)$  a \emph{transformed} version of the median of the network:
\begin{equation}
 \hat{\mu}_s(t) = T(M_t),
\label{E:mu-s}
\end{equation}
where $M_t=\underset{1 \leq i\leq N}{\text{med}} Z_i(t)$ represents the median of the network, and $T$ denotes a transformation composed of an Anscombe transform and a Wiener filtering \citep{Davenport1987}. 
This filtering is applied in order to clean the data from very high frequencies which can lead to instabilities in the subsequent analysis. 
The generalized Anscombe transform stabilizes the variance~\citep{Murtagh1995, Makitalo2013}. It writes as
\begin{equation}
    A(x)=\frac{2}{\alpha}\sqrt{\alpha x + \frac{3}{8} \alpha^2}\ .
    \label{E:Anscombe}
\end{equation}
It is commonly applied in the literature to gaussianize near-Poissonnian variables. It is needed here, as the Wiener filtering performs better on Gaussian data. 
Similarly to~\cite{Dudok2016}, pp.~14-15, we fix $\alpha=4.2$ in Equation~(\ref{E:Anscombe}). This is the optimal value found for the composite $N_c$. 
 Before applying the Wiener filtering, missing values of the median of the network are imputed using the algorithm described in~\cite{Dudok_gap}.  Only 49 values are imputed, which represents 0.2\% of the total number of values on the period studied.
The Wiener filtering is then applied on the transformed and complete set of median values, and  suppresses the highest frequencies of the signal. 
Finally, the imputed missing values were reset to NaN (\lq not a number') in $\hat \mu_s(t)$.
The block diagram of the procedure is described in Figure~\ref{Fig:filtering}.

Among other tested estimators (based on the mean, the median of the network or a subset of stations), with or without application of $T$, the estimator proposed in Equation~(\ref{E:mu-s}) turns out to be the most robust to outliers.

\subsection{Comparison with space data}
\label{ssec:space-data}

To test the quality of our estimator $\hat{\mu}_{s}$, we compare it with a sunspot number extracted from satellite images of the Sun. 
We expect less variability when $N_s$, $N_g$ and $N_c$ are retrieved from satellite images using automated algorithms
as the rules to count spots and groups are clearly defined. Nevertheless, the measurements are biased by these rules and the most complex cases, e.g. at maxima, most often require either human intervention or a specific procedure in the algorithm. 
In any case, a measure of the \lq true' number of spots and groups does not exist.

As exercise for this comparison, we use the Sunspot Tracking And Recognition Algorithm (STARA) sunspot catalogue \citep{Watson2010}, regrouping observations from May 1996 to October 2010 (solar cycle 23). This number is extracted using an automated detection algorithm from the images obtained by the MDI instrument on SOHO. It has a lower scaling than our estimator for the number of spots, $\hat{\mu}_{N_s}$, as expected since the definition of a spot in the detection algorithm excludes the pores (spots without penumbra).

\begin{figure}[hbt]
	\centering
		\includegraphics[width=10cm, height=7cm]{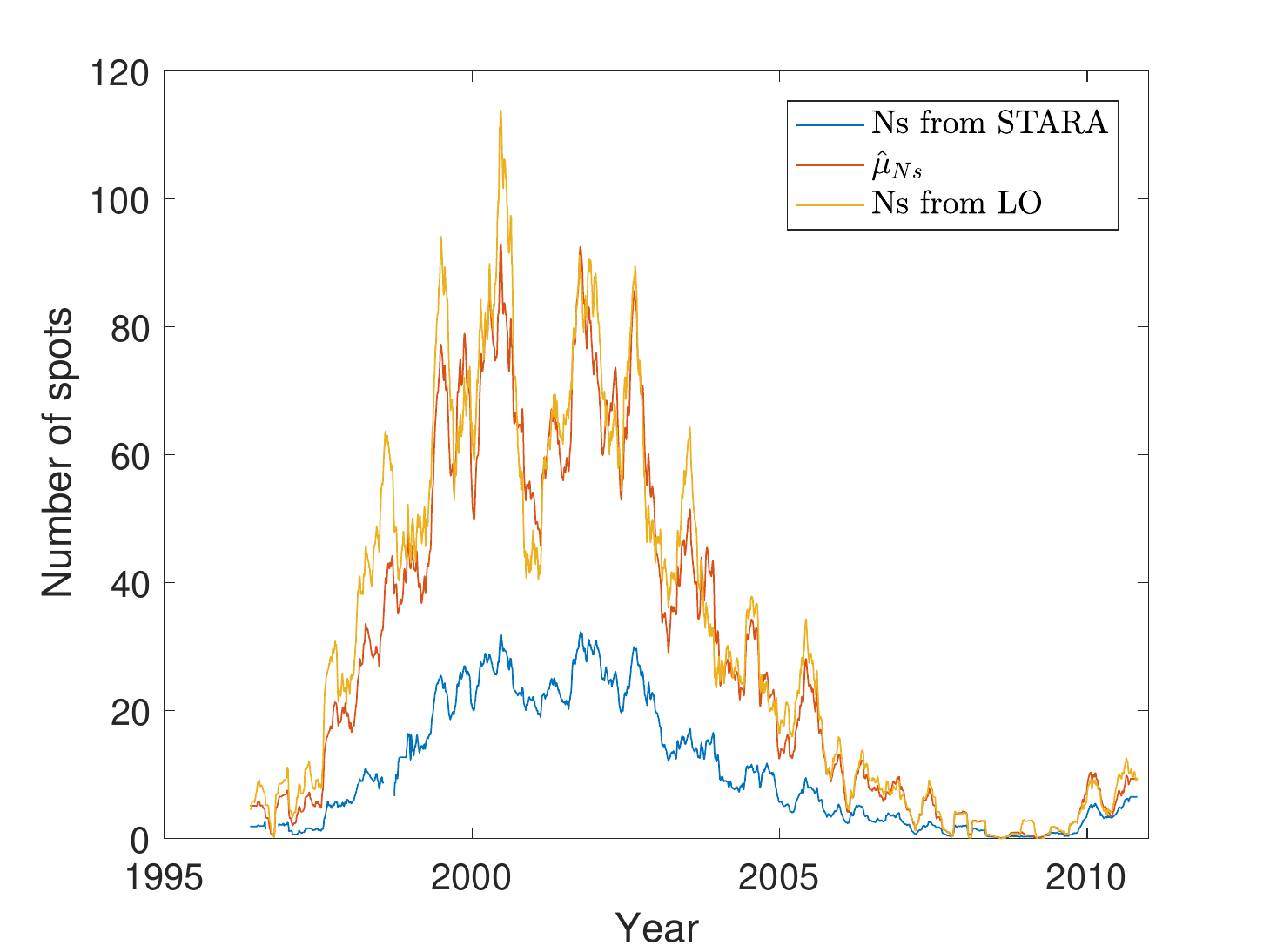}
	\caption{{\small Comparison between the SN obtained from STARA and from our procedures, for the period May 1996 to October 2010. The number of spots obtained from the STARA catalogue is represented in blue, the actual (unprocessed) number of spots observed in Locarno (LO) in yellow, and $\widehat{\mu}_{N_s}$ is plotted in orange. The three quantities shown are averaged over 81 days.}}
	\label{fig:stara}
\end{figure}

We compare three quantities on the period where STARA data are available (1996-2010): $N_s$ (STARA), $\hat \mu_{N_s}$, and $N_s$ as recorded by the Locarno station. These are shown in Figure~\ref{fig:stara}. We test the level of variability by computing the mean value of a moving standard deviation over a window of 81 days. It is equal to $14.07$ for $N_s$ (STARA) re-scaled on $\hat \mu_{N_s}$ ($5.38$ for $N_s$ (STARA) without scaling) against $15.68$ for $\hat \mu_{N_s}$ and $27.13$ for Locarno. As expected, $N_s$ (STARA) experiences less variability than a single station but its variability is comparable to the one of our estimator.  

Nevertheless, satellite images of the  Sun have only been available since 1980, and data extracted from those images cannot be traced back until the 17th century. Gathering space observation during several decades also require the use of different satellites and instruments, as instruments age in space. 
These instruments need calibrations that create additional inaccuracies to the extracted numbers. We thus conclude that $ \widehat{\mu}_s(t)$ is a more robust estimator of the solar activity, and it will be used as a proxy for $s(t)$ in the sequel.

\subsection{Solar component for Ns and Ng}

We present here the statistical modelling of the number of spots $N_s$ and  the number of groups $N_g$. We do this \emph{separately} for each component since their physical origin are driven by different phenomena: the groups convey information about the dynamo-generated magnetic field in the solar interior whereas the emergence of individual spots would rather come from fragmented surface flux and agglomeration of small magnetic fields,~\citep{Thomas2008}. Together, the analysis of the spots and groups helps us to better understand the composite $N_c$ and the solar activity which is not satisfactorily described by only one of the two numbers~\citep{Dudok2016}.  In the remainder of this paper, we define a specific notation for the generic $\hat \mu_s(t)$ from Equation~(\ref{E:mu-s}): $\hat \mu_{N_s}(t)$ for the number of spots, $\hat \mu_{N_g}(t)$ for the number of groups and $\hat \mu_{N_c}(t)$ for the composite. 

\begin{figure}[htb]
\begin{minipage}[b]{.49\linewidth}
  \centering
  \centerline{\includegraphics[width=6.5cm, height=5cm]{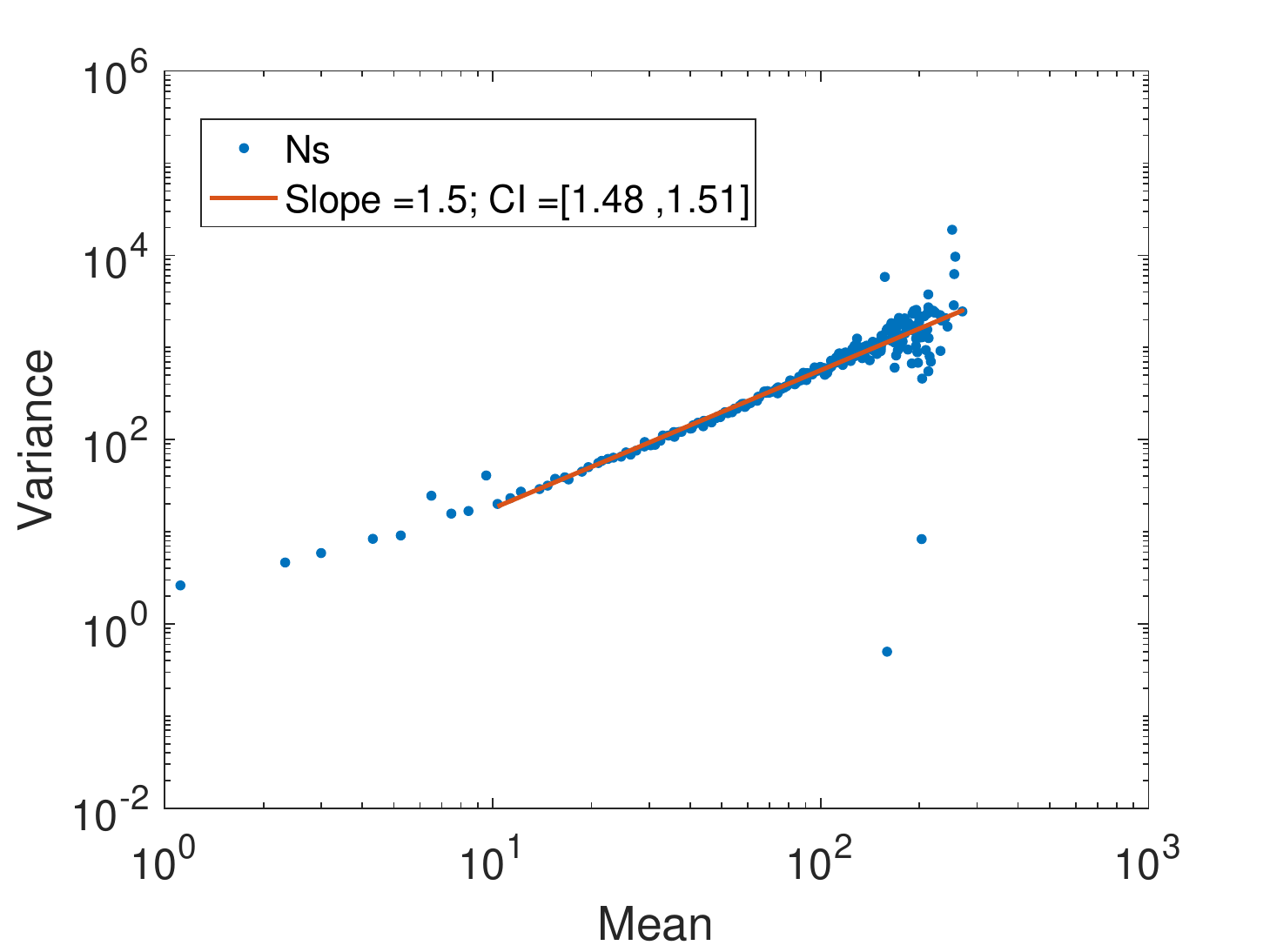}}
\end{minipage}
\hfill
\begin{minipage}[b]{0.49\linewidth}
  \centering
  \centerline{\includegraphics[width=6.5cm, height=5cm]{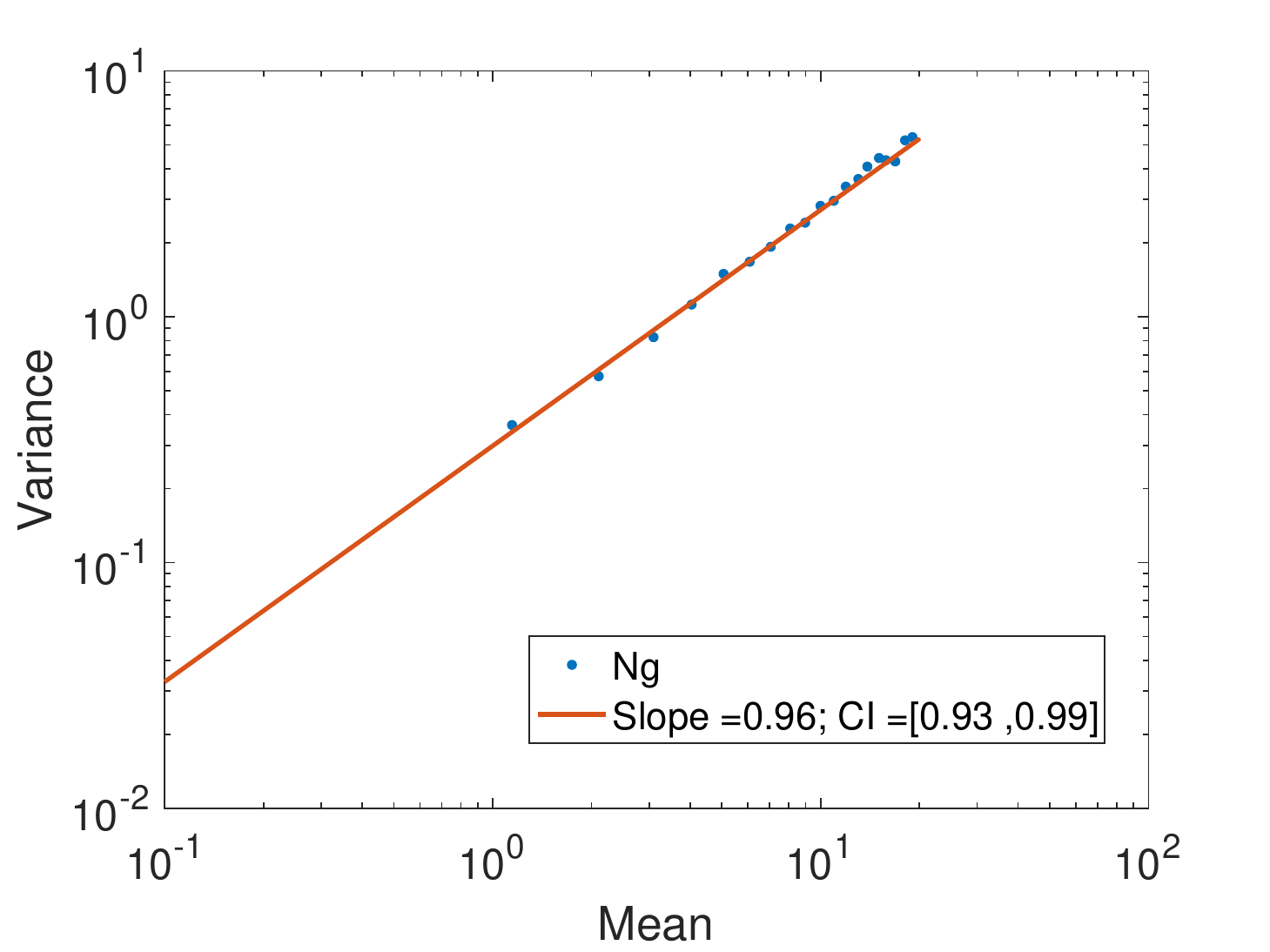}}
\end{minipage}
\caption{\small{Estimation of the conditional mean-variance relationship for $N_s$ (left) and $N_g$ (right). The red line is a linear fit of the points (shown on a log-log scale), starting at $N_s > 9$ and $N_g > 0$, respectively. In both plots, the legend shows the value of the fitted slope together  with its confidence interval at 95\%. 
The value of the intercept is $-1.21$ for $N_g$ and $-0.57$ for $N_s$.
The same fit starting at $N_s > 0$ (not shown here) leads to a slope of $1.25$ and $CI_{95\%}=[1.23, 1.28]$.}}
\label{fig:scatter}
\end{figure}

\begin{figure}[htb]
\begin{minipage}[b]{.49\linewidth}
  \centering
  \centerline{\includegraphics[width=6.5cm, height=4.8cm]{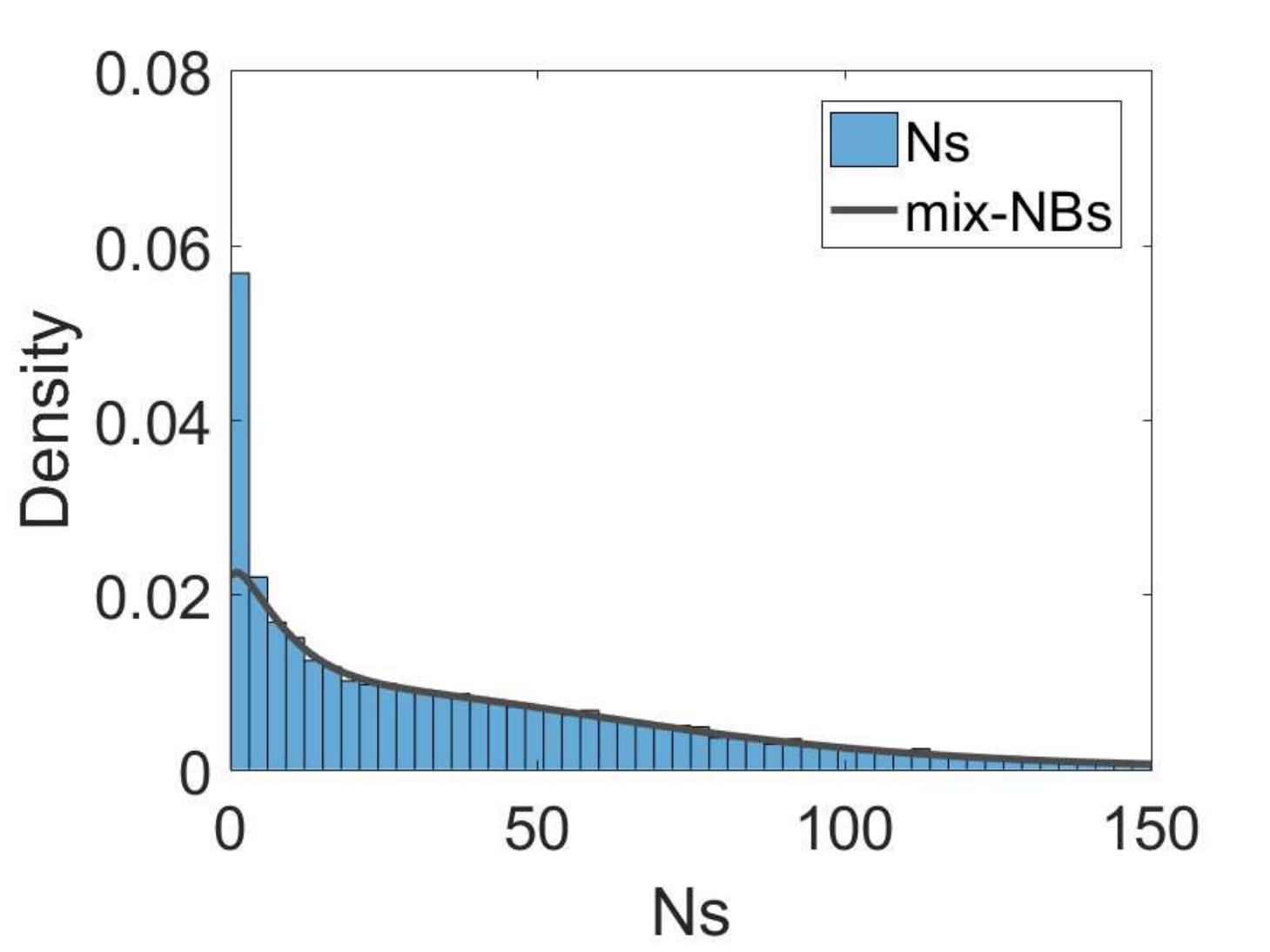}}
\end{minipage}
\hfill
\begin{minipage}[b]{0.49\linewidth}
  \centering
  \centerline{\includegraphics[width=6.5cm, height=4.8cm]{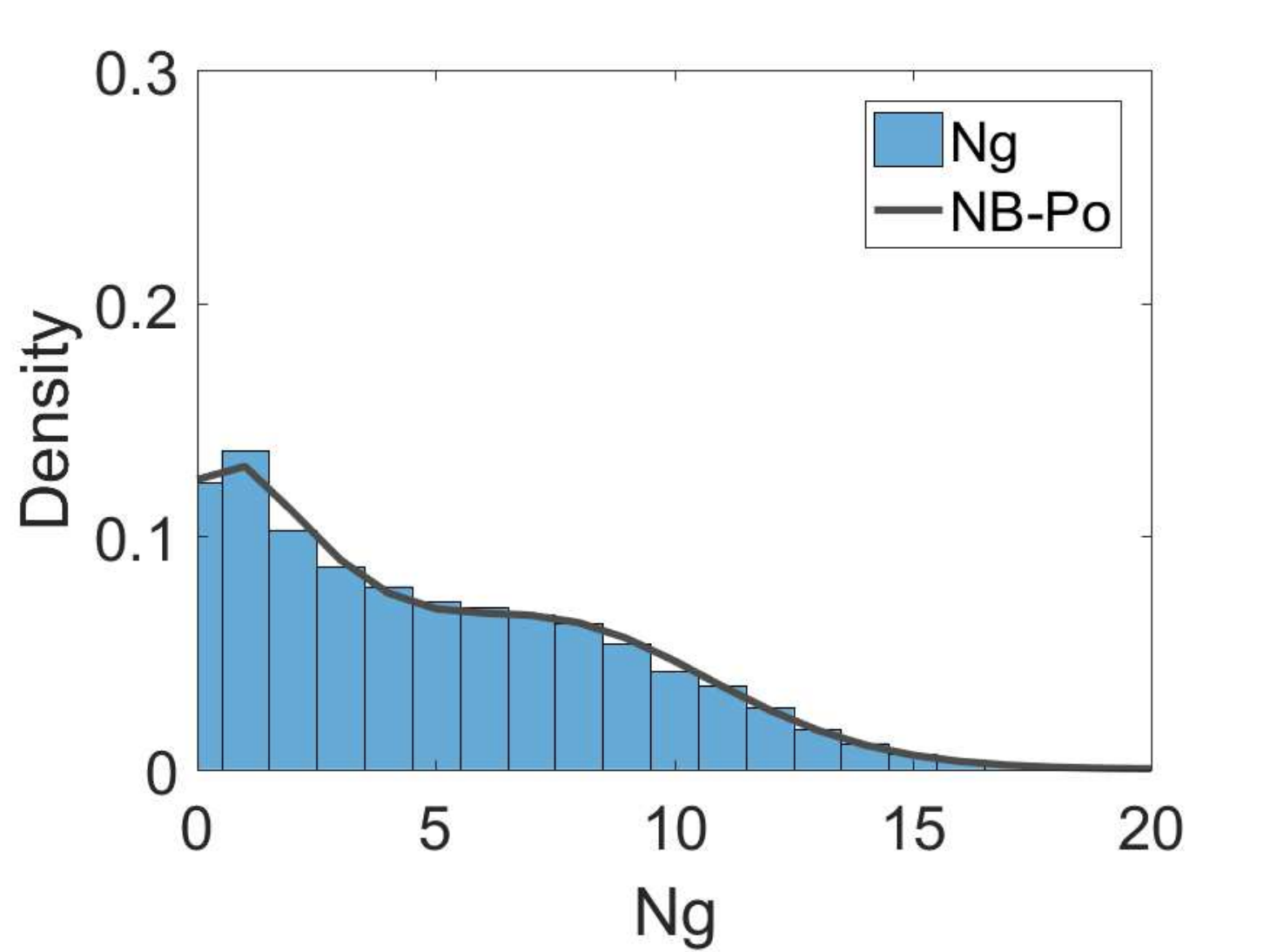}}
\end{minipage}
\caption{\small{
(Left) Histogram of $\widehat{\mu}_{N_s}(t)$ values, computed with a bandwdith (binning) equal to 3, and estimated density for non-zero values of $\widehat{\mu}_{N_s}(t)$ (shown in black line). The complete density is modelled by a ZA mixture of generalized NBs. For the zero values, the MLE value of the Bernouilli parameter is equal to
$b=0.1$.  For non-zero values, the MLE values of the parameters in~Equation(\ref{E:ZANB}) are: $r_1=1.25$, $p_1=0.11$, $r_2=2.39$, $p_2=0.04$ and $w_1=0.32$. 
(Right) Histogram of $\widehat \mu_{N_g}(t)$ values, computed with a bandwidth equal to one, and corresponding density fitted by MLE (shown in black line). The density is modelled by a mixture of 
 an NB and a Poisson distributions as defined in Equation~(\ref{E:NBP}). The fitted parameter values are:  $\mu_2=8.62$, $r_1=1.65$, $p_1=0.37$, and $w_1=0.36$. }}
\label{fig:mus}
\end{figure}

The authors in~\cite{Dudok2016} showed that the number of spots and groups experience more over-dispersion than actual Poisson variables.  In order to estimate how far the distribution of the \lq true' $s(t)$ departs from a Poisson distribution, 
we regress the conditional variance $\var(Z_i(t)|\hat \mu_{s}(t) = \mu)$ 
versus  the conditional mean $\EE(Z_i(t)|\hat \mu_{s}(t) = \mu)$
by OLS, see Figure~\ref{fig:scatter}. Whereas in a Poisson context, the slope of the fit should be close to one, for  $N_s>10$, 
our fit shows a slope of $1.5$, with confidence interval (CI) $CI_{95\%}=[1.48, 1.51]$. This points to overdispersion and the need for a generalization of a Poisson PDF. On the contrary, the same plot for $N_g>0$ displays a slope of $0.96$, with $CI_{95\%}=[0.93, 0.99]$, confirming the validity of a Poisson process assumption. Note that the values $<11$ are excluded from the fit of $N_s$, as they seem to indicate a different regime. This change in the alignment may indicate the presence of a multi-modal distribution, see Figure~\ref{fig:mus}. 

Count data with over-dispersion are widely modelled by the negative binomial (NB) distribution in the literature \citep{count_data, Rodriguez2013} or by its generalization \citep{Jain1970}: 
\begin{equation}
    g(x,r,p)=\frac{\Gamma(r+x)}{\Gamma(r)\Gamma(x+1)}p^r q^x,
    \label{E:NB}
\end{equation}
where $r>0$, $p \in (0,1)$, $q=(1-p)$ and $\Gamma$ is the gamma function. 

A visual inspection of the histogram of estimated values $\widehat{\mu}_{N_s}(t)$ in Figure~\ref{fig:mus}(Left) reveals a local maxima in the distribution around $20-40$. We refer to these local maxima as \emph{modes} in the remainder of the article. The underlying density of $\hat \mu_{N_s}(t)$ may thus be multi-modal, as suspected from Figure~\ref{fig:scatter}(Left). Such PDFs are classically modelled by a mixture model.  As the density shows a typical excess of zeros as well, it requires the use of a ZA distribution defined in Equation~(\ref{E:hurdle}). 
We thus fit the complete PDF of the estimated number of spots, $\hat \mu_{N_s}(t)$, by a ZA mixture of generalized NB distributions. The density at zero, $f_0(x)$, is represented by a Bernoulli distribution, whereas the density outside zero, $f_1(x)$ in Equation~(\ref{E:hurdle}), is identified by a mixture of NB distributions:
\begin{equation}
   f_1(x,r_1,r_2,p_1,p_2)= w_1 g_1(x,r_1,p_1) + (1-w_1)g_2(x,r_2,p_2),
    \label{E:ZANB}
\end{equation}
where $g_1,\, g_2$ are NB densities and $w_1$ is the mixture weight. 

Similarly, the histogram of  $\hat \mu_{N_g}(t)$ exhibits a clear excess in the range $1-3$ compared to a Poisson-like distribution centred between 5 and 8. The PDF of $\widehat{\mu}_{N_g}(t)$ shows thus two modes: one around $1-3$, and  one around $5-8$.  Such PDF may be modelled by a mixture of a NB and a Poisson distribution: 
\begin{equation}
  f(x,r_1,p_1,\mu_2)=w_1 g(x,r_1,p_1) + (1-w_1) \frac{\mu_2^x}{x!}e^{\mu_2},
    \label{E:NBP}
\end{equation}
where $\mu_2>0$ and, as above, $w_1$ is the mixture weight.

The fit of these parametric densities are shown in Figure~\ref{fig:mus} by a black line superimposed on the histograms. All  the fits in this paper article are computed using the maximum likelihood estimation (MLE).  The nature of the different modes in the PDF of $\hat \mu_{N_s}$ and $\hat \mu_{N_g}$will be discussed in Section~\ref{sec:corr}.

\subsection{Solar component for $N_c$}

\begin{figure}[!hbt]
\begin{minipage}[b]{.49\linewidth}
  \centering
  \centerline{\includegraphics[width=6.5cm, height=4.8cm]{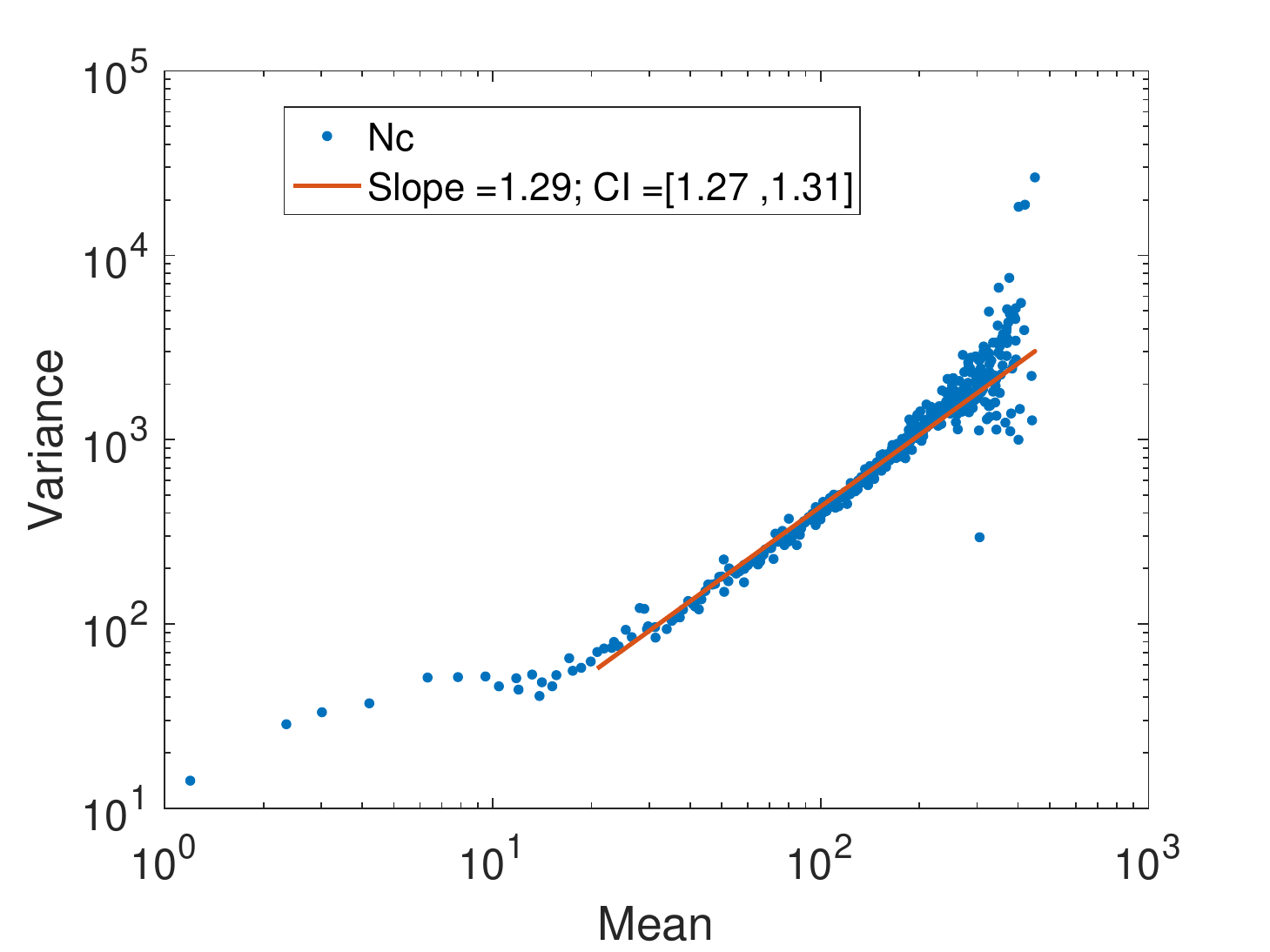}}
\end{minipage}
\hfill
\begin{minipage}[b]{0.49\linewidth}
  \centering
  \centerline{\includegraphics[width=6.5cm, height=4.8cm]{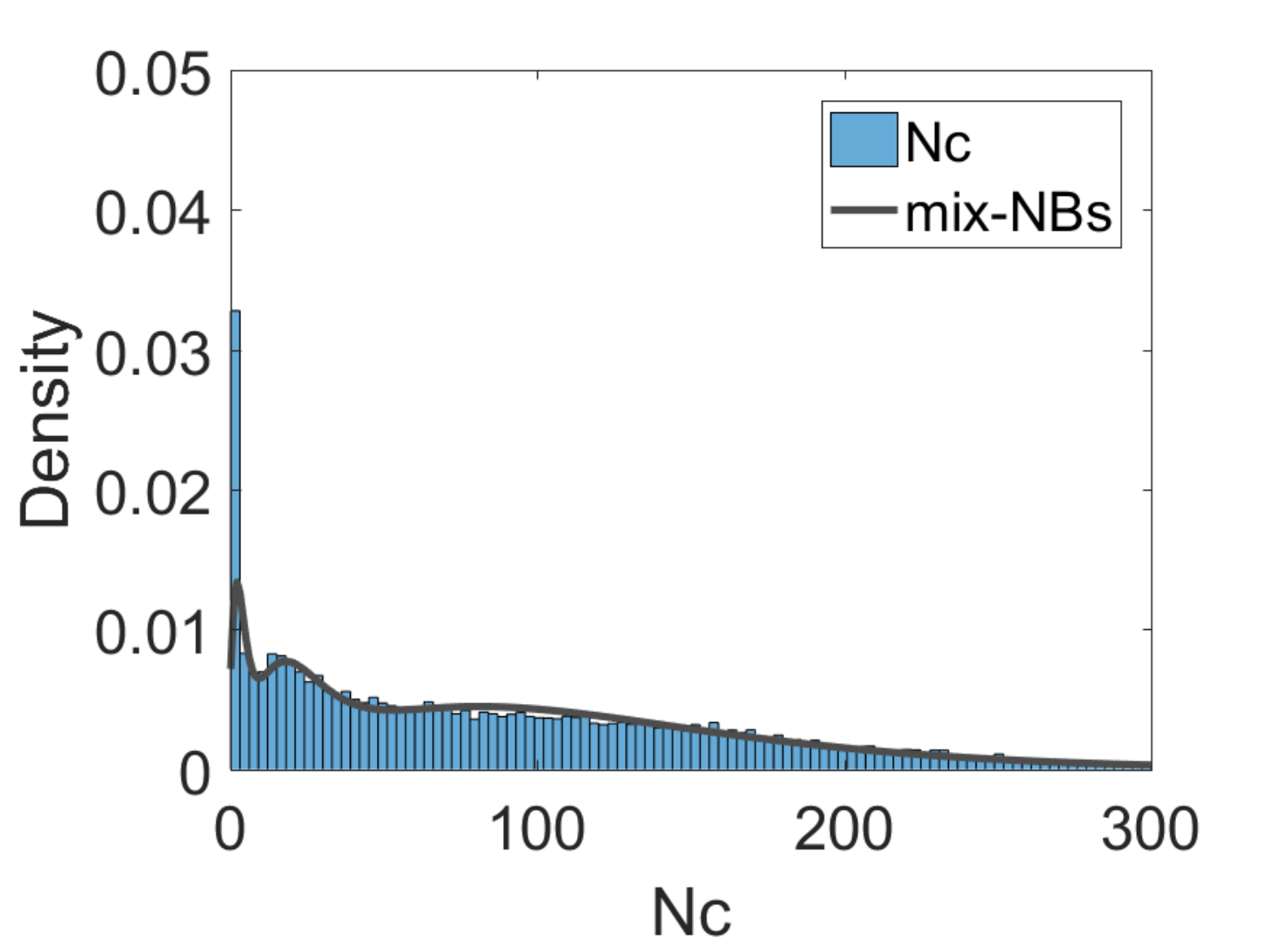}}
\end{minipage}
\caption{\small{
(Left) Conditional mean-variance relationship for $N_c$, shown on a log-log scale.
(Right) Histogram of $\hat \mu_{N_c}(t)$ values, with a binning equal to $bw=3$. The estimated density outside of zeros values is shown by a black line. It is modelled as a mixture of three NB distributions,   see Equation~(\ref{E:3NB}), with MLE parameter values equal to: $r_1=3.18$, $p_1=0.48$, $r_2=4.02$, $p_2=0.15$, $r_3=3.05$, $p_3=0.02$, $w_1=0.08$ and $w_2=0.19$. The Bernoulli parameter of the density $f_0$ at zero is equal to $b=0.07$.}} 
\label{fig:mus-sn}
\end{figure}

We now use Equation~(\ref{E:mu-s}) to estimate the  $\mu_{N_c}$, the \lq true' value of the composite $N_c=N_s+10N_g$. 
Again looking at the conditional mean-variance relationship, we observe in Figure~\ref{fig:mus-sn}(Left) an over-dispersion with a slope of $1.29$ and $CI_{95\%}=[1.27, 1.31]$ for $N_c>20$. As a compound of both quantities, $N_c$ experiences less over-dispersion than $N_s$  and more than $N_g$. 
A visual inspection of the histogram of $\hat \mu_{N_c}(t)$ values in Figure~\ref{fig:mus-sn}(Right) indicates an excess of zeros and several modes, most probably coming from the modes observed in the PDFs of $\hat \mu_{N_g}$ and $\hat \mu_{N_s}$. 
We find thus appropriate to approximate the density of $\hat \mu_{N_c}(t)$  by a ZA mixture of three NB distributions, where the density outside zero values, $f_1(x)$ in Equation~(\ref{E:hurdle}), is identified with:
\begin{equation}
    f_1(x,r_1,...,r_3,p_1,..,p_3)=\sum_{i=1}^{3} w_i g_i(x,r_i,p_i), 
    \label{E:3NB}
\end{equation}
where $w_i$ are the mixture weights and $\sum_{i=1}^{3} w_i=1$.  The fit of $f_1$ is represented in Figure~\ref{fig:mus-sn}(Right) by a black line. \\

A statistical analysis (not presented here) shows that the distribution of the ISN is statistically close to the distribution of $\hat \mu_{N_c}$. The uncertainty analysis for $\hat \mu_{N_c}$, presented in the remainder of the article, remains thus valid for the ISN. 

\subsection{Conditional Correlation}
\label{sec:corr}

Due to the physical nature of the data, the local maxima for the densities
of $\widehat{\mu}_{N_s}$ and $\widehat{\mu}_{N_g}$ are not independent. We therefore look at the conditional correlation $Corr(N_s,N_g|\widehat{\mu}_{N_c}=s)$ 
with the goal to better understand the nature of the modes observed in these two densities,  and thus also in the density of $\hat \mu_{N_c}$. 

\begin{figure}[!htb]
	\centering
	\begin{subfigure}{0.32\textwidth}
		\centering
		\includegraphics[width=5.5cm, height=4cm]{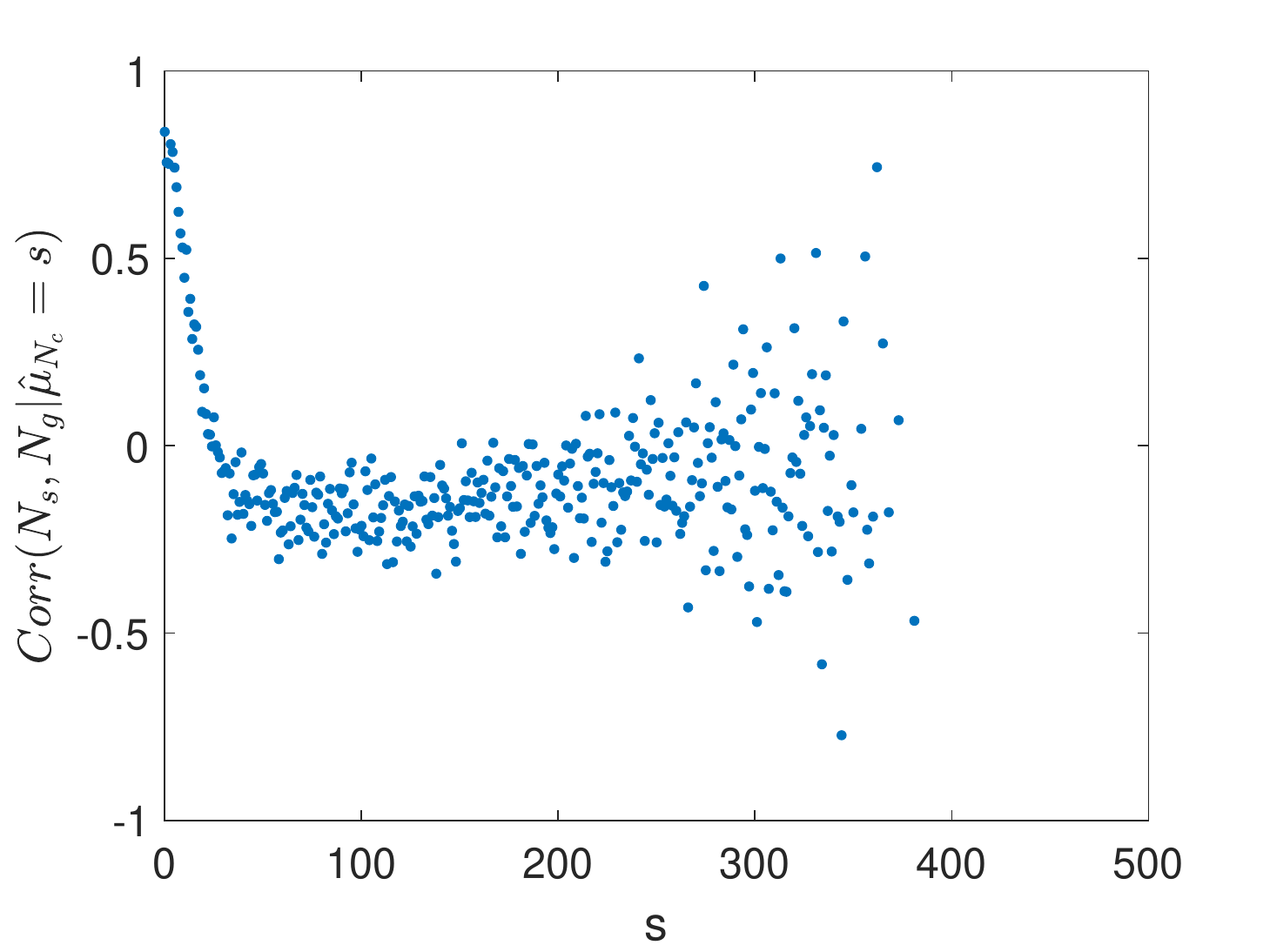}
		\caption{}
		\label{fig:corr}
	\end{subfigure}
	\begin{subfigure}{0.32\textwidth}
		\centering
		\includegraphics[width=5.5cm, height=4cm]{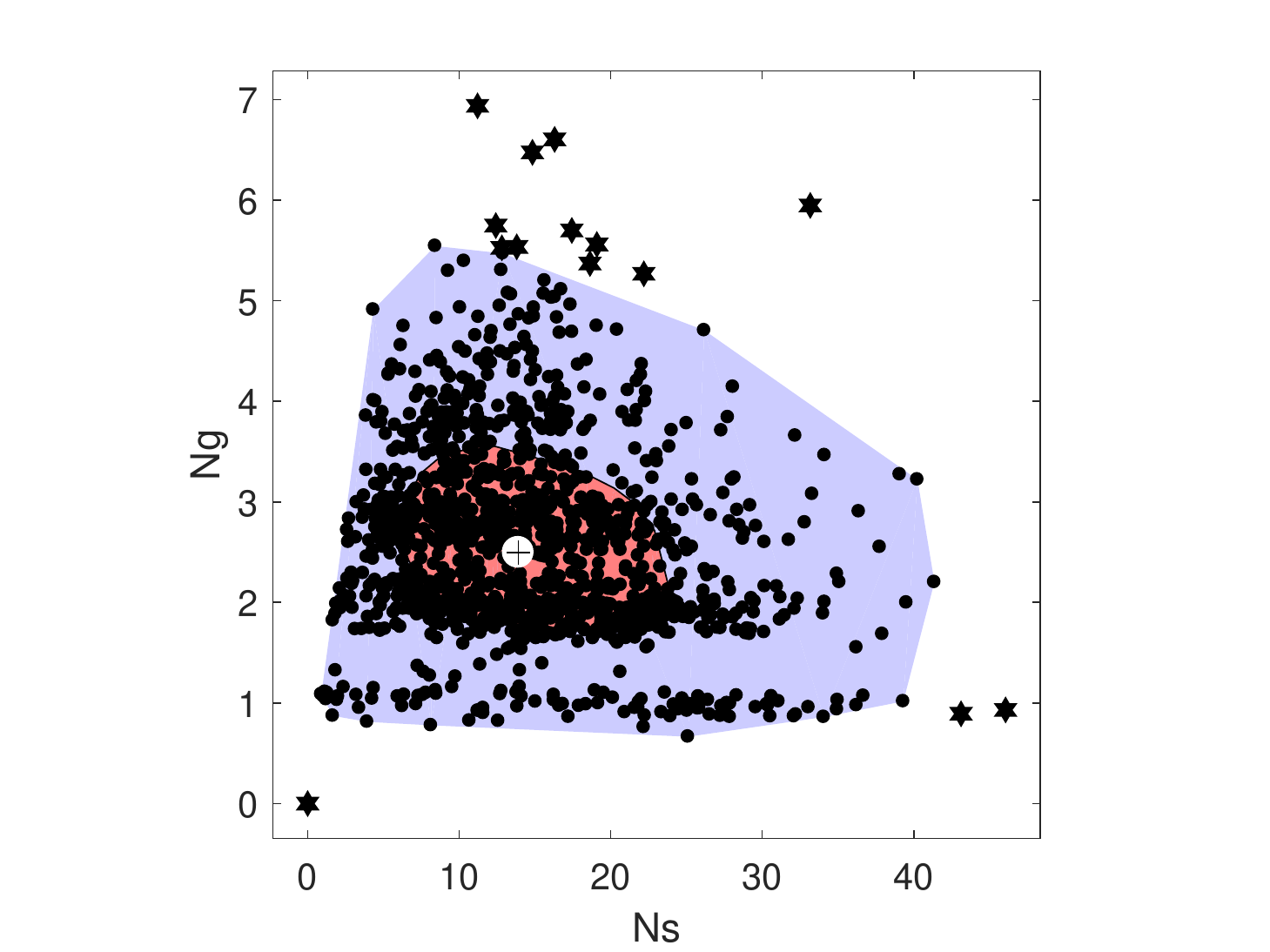}
		\caption{}
		\label{fig:bag40}
	\end{subfigure}
	\begin{subfigure}{0.32\textwidth}
		\centering
		\includegraphics[width=5.5cm, height=4cm]{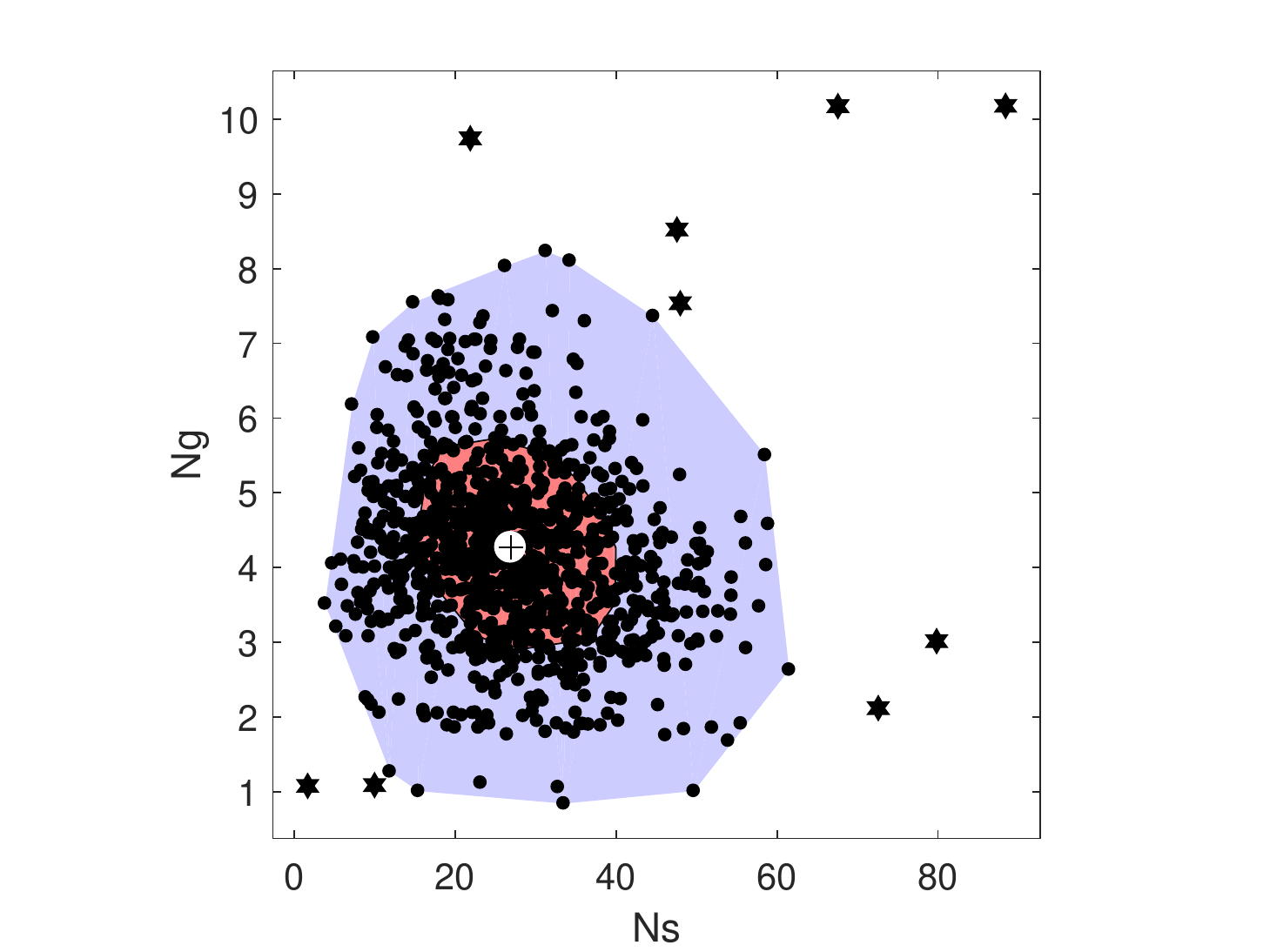}
		\caption{}
		\label{fig:bag80}
	\end{subfigure}
	\caption{\small{Conditional correlation of $N_s$ and $N_g$: $Corr(N_s,N_g|\hat \mu_{N_c}=s)$ for $s \in [0,400]$ (a). 	Bivariate boxplot (also called \lq bagplot') of $N_s$ and $N_g$ when $\hat \mu_{N_c}=40$ (b) and $\hat \mu_{N_c}=70$ (c).
The white cross represents the depth median \citep{Rousseuw2012}. The bag contains 50\% of the observations and it is represented by a polygon in red. The fence (not represented) is obtained by inflating the bag by a factor three. The observations that are outside of the bag but inside of the fence are indicated by a light grey loop. Outliers are represented by a black star. The correlation is indicated by the orientation of the bag.}}
\end{figure}

Figure~\ref{fig:corr} shows the conditional correlation for different values of $\hat \mu_{N_c}$ between 0 and 400. Note that even when $\hat{\mu}_{N_c}=s$, the value of the composite $N_s+10 N_g$ for a particular station may be larger (resp. smaller) than $s$. Our analysis highlights  three regimes of activity:
\begin{description}
\item[Minima]  $\hat \mu_{N_c} \in [0,11]$. Here, the number of spots and groups oscillates between 0 or 1. As the number of spots equals exactly the number of groups, the correlation is high.
\item[Medium activity] $\hat \mu_{N_c} \in [12,60]$. The correlation progressively decreases, because the number of spots increases faster than the number of groups, and then stabilizes. This regime is characterized by the development of smaller spots without penumbra or with a small penumbra.
Figure~\ref{fig:bag40} shows the bivariate boxplot of $N_s$ and $N_g$ when $\hat \mu_{N_c}=40$.
For $N_g=1$ or $N_g=2$, we observe values of $N_s$ as high as 40. 
We clearly observe groups containing a large number of spots as well as groups, composed of fewer spots, that appear progressively as the penumbra grows and that indicate a transition toward groups with fewer but larger spots. The effect of this transition from small to larger spots is observed in Figure 5 from~\cite{all_corrections}. 
\item[High activity] $\hat \mu_{N_c}>60$.  
Figure~\ref{fig:bag80} shows the bivariate boxplot of $N_s$ and $N_g$ when $\hat \mu_{N_c}=70$. The plot has a potato-shape around ($N_s=30$, $N_g=4$). We now observe all kinds of groups. The correlation between groups and spots slightly increases as the number of groups begins to grow as well. 
\end{description}

The \emph{medium} and \emph{high} regimes are reflected in the estimated densities of $\widehat{\mu}_{N_s}$ and $\widehat{\mu}_{N_g}$ in Figure~\ref{fig:mus}.
The first mode of $\hat \mu_{N_g}$ ranging from 1-3 corresponds to the \emph{medium} regime while the second from 5-8 reflects the \emph{high} regime. The two distinct regimes provide another justification for the use of a multi-modal distribution to characterize the PDF of $\hat \mu_{N_g}$. Similarly, there is also a mode in the distribution of $\widehat{\mu}_{N_s}$ around 20-40 that comes from the transition between the \emph{medium} and the \emph{high} regime. The mode is correctly represented by a mixture model. The study of the conditional correlation constitutes the first step towards retrieving the distribution of $N_c$ from its composites $N_s$ and $N_g$. However, this task is challenging and goes beyond the scope of the article because: (1) the distributions of $N_s$ and $N_g$ are complex mixtures and (2) the number of spots is non-trivially correlated to the number of groups.

\section{DISTRIBUTION OF ERRORS}
\label{sec:errors}

We are now in a position to analyze the error distribution in sunspot counts, the modelling of the distributions of $\epsilon_3$, $\epsilon_1$ and $\epsilon_2$. To do so, we separate minima from non-mimima regimes. We also consider two time-scales: short-term periods, that is, time-scales smaller than one solar rotation (27 days), and long-term periods. Section~\ref{S:error_minima} estimates error at solar minima, i.e. when $s(t)=0$. Section~\ref{S:error_short-term} analyzes short-term variability of the pre-processed observations when $s(t)>0$. For the study of long-term error in Section~\ref{S:error_long-term}, we use raw data that did not undergo any pre-processing, in order to be able to detect sudden jumps and/or large-drifts in the time series. 
The correct time-scale for the long-term period is also determined in this section, based on a statistically-driven procedure. Finally, Section~\ref{S:error_comparing} 
compares the characteristics of the different stations based on the error analysis.  

\subsection{Error at minima}
\label{S:error_minima}
The study of solar minima periods is complex as the data show a large variability and dichotomy.
Observed values of the error at minima, $\epsilon_3$, are defined as counts made by the stations when the proxy for $s(t)$, defined in Equation~(\ref{E:mu-s}), is equal to zero:
\begin{equation}
\hat\epsilon_3(i,t) = Z_i(t) \ \text{when} \ \hat \mu_s(t)=0,
\label{E:e3}
\end{equation}
where $Z_i(t)$ corresponds either to $N_s$, $N_g$ or $N_c$, and where the generic $\widehat{\mu}_s(t)$ has to be replaced by $\widehat{\mu}_{N_s}(t)$ for $N_s$, $\widehat{ \mu}_{N_g}(t)$ for $N_g$, and $\widehat{\mu}_{N_c}(t)$ for $N_c$. 

\begin{figure}[bt!]
	\centering
		\includegraphics[scale=0.4]{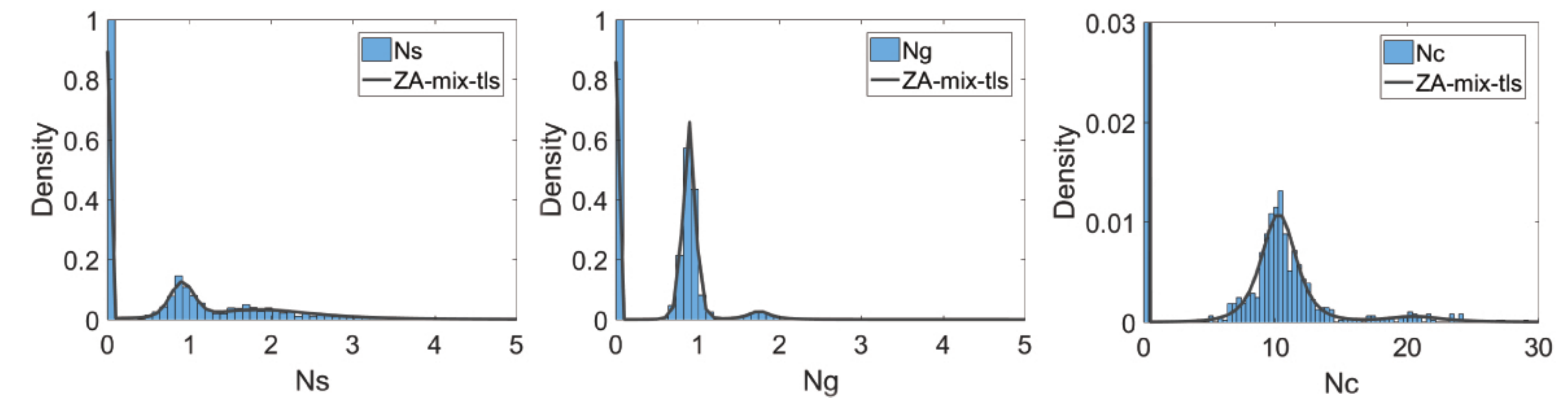}
\caption{\small{Truncated histograms of $\hat \epsilon_3$ for $N_s$ (a), $N_g$ (b) and $N_c$ (c).  The continuous line shows the fits using a ZA-mixture of $t$-LS distributions, defined in Equation~(\ref{E:mixtls}).  The values of the Bernouilli parameter in  Equation~(\ref{E:hurdle}) are equal to: (a) b=0.9, (b) b = 0.86, and (c) b= 0.96
They represent the proportion of \lq true' zeros. The parameter values for the  $t$-LS fit are:
(a) for $N_s$: $\mu_1=0.91$, $\sigma_1=0.14$, $\nu_1=31.16$, $\mu_2=1.85$, $\sigma_2=0.71$, $\nu_2=2.09$, $w_1=0.6$;  (b)
for $N_g$: $\mu_1=0.89$, $\sigma_1=0.07$, $\nu_1=6.89$, $\mu_2=1.75$, $\sigma_2=0.14$, $\nu_2=1.33$, $w_1=0.09$; (c) for $N_c$: 
 $\mu_1=10.24$, $\sigma_1=1.37$, $\nu_1=3.89$, $\mu_2=20.57$, $\sigma_2=2.33$, $\nu_2=1.93$, $w_1=0.08$.
 The bin-width ($bw=0.0917$) is the same for the histograms of both $N_s$ and $N_g$. It is related to the sample size and the data-range of $N_s$ by a simple rule proposed by Scott~\citep{Scott1979}. The bin-width of the histogram of $N_c$ in (c) is equal to $bw=0.4192$ and is also computed by Scott's rule. 
Note that the right figure is enlarged: the value at zero is $0.96$ and not $0.03$.} }
\label{fig:e3}
\end{figure}

A visual inspection of the histogram of $N_s$ (resp.  $N_g$) in Figure~\ref{fig:e3}(Left) (resp. Figure~\ref{fig:e3}(Center)) shows an important amount of \lq true' zeros together with two modes around one  and two. Similar modes occur around 11 and 22 in the distribution of $N_c$ in Figure~\ref{fig:e3}(Right), as expected. 
 These modes represent short-duration sunspots. Due to the non-simultaneity of the observations between stations, the proxy for $s(t)$ might be equal to zero even if some spots appear shortly (from several minutes to several hours) on the Sun. These modes can be represented by a $t$-Location-Scale ($t$-LS) distribution, which is a generalization of the Student t-distribution.  This distribution has three parameters to accomodate for asymmetry and heavy tails: the location $\mu$, scale $\sigma>0$, and shape $\nu>0$, see~\cite{tLS,distributions}. Its PDF is defined as: 
\begin{equation}
g(x,\mu,\sigma,\nu)_{t-LS} = \frac{\Gamma(\frac{\nu+1}{2})}{\sigma \sqrt{\nu \pi} \Gamma(\frac{\nu}{2})} \left( \frac{\nu + \frac{(x-\mu)^2}{\sigma^2}}{\nu} \right)^{-(\frac{\nu +1}{2}) }.
\label{E:tls}
\end{equation}
The large proportion of zeros values for $\hat \epsilon_3$ requires the use of a ZA model as in Equation~(\ref{E:hurdle}). We choose a ZA mixture of $t$-LS for the complete distribution of $\hat \epsilon_3$. The density outside of zero, $f_1(x)$ in Equation~(\ref{E:hurdle}), is  thus identified by such a mixture of $t$-LS distributions: 
\begin{equation}
f_1(x,\mu_1,\sigma_1,\nu_1,\mu_2,\sigma_2,\nu_2) =  w_1g(x,\mu_1,\sigma_1,\nu_1)_{t-LS} + (1-w_1) g(x,\mu_2,\sigma_2,\nu_2)_{t-LS}
\label{E:mixtls}
\end{equation}
where, as before, $w_1$ is the mixture weight. The histograms and fitted distributions for $\widehat{\epsilon}_3$ are shown in Figure~\ref{fig:e3}. The visual closeness between the histogram and the fitted distribution was used as a criterion to select the best PDF among a few intuitive candidates, while the parameters of the distribution are estimated via MLE. \\

In the previous figures, where the error at minima is represented for all stations combined, outliers defined as $\hat \epsilon_3(i,t)>2$ are not visible for $N_g$ and $N_s$. A separate analysis (not presented here) shows that the percentage of outliers in each station are low (inferior to 0.5\% for $N_s$). Some stations also observed a high maximal value at minima (e.g. a value of 35 was recorded in QU (Quezon, Philippines) for $N_s$). This extreme value for a minima may correspond to a transcription error that might be verified in the future, before being encoded in the SILSO database.

\subsection{Short-term variability}
\label{S:error_short-term}

When the proxy for $s(t)$, defined in Equation~(\ref{E:mu-s}), is different from zero, the short-term error $\widetilde{\epsilon}$ may be estimated using: 
\begin{equation}
\widehat{ \widetilde{\epsilon}}(i,t) = \frac{Z_i(t)}{\hat \mu_s(t)}~ \ \text{when} \ \hat \mu_s(t)>0.
\label{E:e1}
\end{equation}
 To select the best distribution we proceed as follows. Different densities are fitted to the values of $\widehat{\widetilde{\epsilon}}$, outside of zero, using MLE.\footnote{We use the function \lq'allfitdist.m', last modified in 2012, in Matlab R2016b.} Then, the AIC criterion is used to choose the best PDF, which in this case is a $t$-LS distribution.
 
\begin{figure}[!ht]
	\centering
		\includegraphics[scale=0.38]{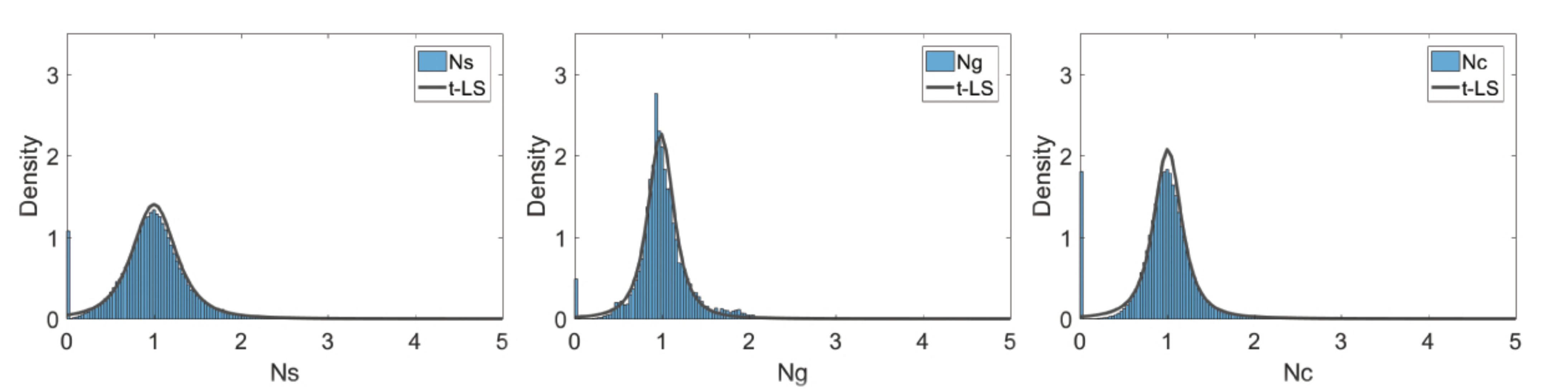}
\caption{\small{Histograms of $\hat{\widetilde{\epsilon}}$ for $N_s$ (a), $N_g$ (b) and $N_c$ (c).  The continuous line shows the fits using a $t$-LS distribution defined in Equation~(\ref{E:tls}).  The values of the Bernoulli parameter in  Equation~(\ref{E:hurdle}) are equal to: (a) $b=0.04$ , (b)  $b=0.02$, (c) $b=0.06$. They represent the proportion of false \lq zeros', i.e. stations reporting no sunspot where there are some. The parameter values for the  $t$-LS fit are: (a) for $N_s$: $\mu=1$, $\sigma=0.26$, $\nu=2.8$; (b) for $N_g$: $\mu=0.99$, $\sigma=0.16$, $\nu=2.33$; (c) for $N_c$:  $\mu=1.01$, $\sigma=0.17$, $\nu=2.12$. The bin-widths $bw$ of the histograms are computed using Scott's rule. For $N_s$ and $N_g$ they are the same ($bw=0.0328$ ) and for $N_c$ it is equal to $bw=0.0433$.}}
\label{fig:e1}
\end{figure}

As we observe an excess of zero, we need a ZA $t$-LS distribution to represent the complete distribution of $\widehat{ \widetilde{\epsilon}}$.
Figure~\ref{fig:e1} shows the histogram as well as the fitted PDF of $\widehat{ \widetilde{\epsilon}}$ outside of zero. For the latter, the mean is close to 1, indicating that on average the stations are aligned with $\hat \mu_s(t)$. The histogram exhibits a probability mass at zero representative of \lq false' zeros, that is, of stations that do not observe any sunspot when there are actually some on the Sun. The histogram also shows a tail on the right hand side, caused by outliers. This asymetry requires a $t$-LS rather than a Gaussian distribution to be fitted.

\begin{figure}[hbt]
	\centering
		\includegraphics[width=14cm, height=7cm]{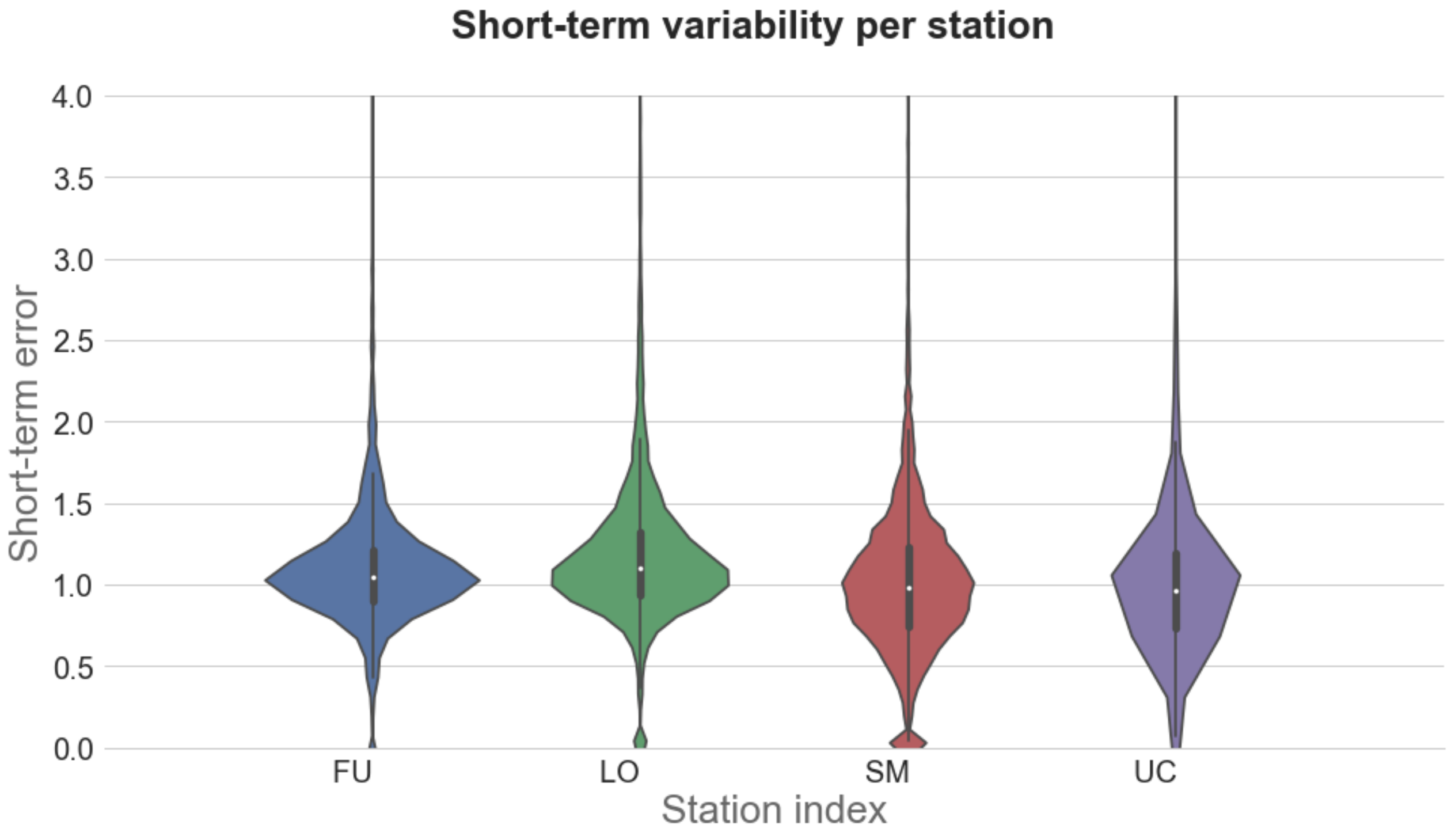}
	\caption{{\small Truncated violin plots of the estimated short-term variability $\widehat{\widetilde{\epsilon}}$ for $N_s$ in four stations (FU, LO, SM and UC). 
	A violin plot \citep{Hintze1998} combines a vertical box-plot with a smoothed histogram represented symmetrically to the left and right of the box. 
	The white dot in the centre of the violin locates the mean of the distribution. The thick grey bar shows the interquartile range and the thin grey bar depicts the interdecile range. 
	The bin-width ($bw=0.05$) is the same for all stations and is computed with Scott's rule.}}
	\label{fig:e1-station}
\end{figure}

The violin plots of four different stations are shown in Figure~\ref{fig:e1-station} for the number of spots $N_s$, where the differences between the stations are the most visible.  The mean of the Locarno station (LO), the current reference of the network, is slightly higher than the three other means (and higher than the means of all other stations), around $1.19$. This results from its particular way of counting: large spots (with penumbra) count for more than small spots without penumbra. 

Another characteristic feature is how the error is distributed around the mean. A violin plot may be seen as a PDF with the x-axis of the density drawn along the vertical line of the boxplot. For example, the PDF of the short-term error of LO is concentrated around the mean but the entire distribution is shifted upward unlike the PDF of Uccle (UC) which has much lower values.   
UC is a professional observatory. Different observers record from one week to another the number of spots, groups and composite on the Sun. As their experience and methodology slightly change, the shift of observers probably increases the short-term variability of the station. Usually a team of observers experience more variability than a single person, like in FU (Fujimori, Japan). This station has a remarkable short-term stability. \\
Similarly, the San Miguel (SM) station shows the typical shape of a professional observatory. On the other hand, the LO station shows an $\widehat{\widetilde{\epsilon}}$ distribution almost characteristic of a single observer: that is because until recently LO had one dominant main observer.  

\subsection{Long-term variability}
\label{S:error_long-term}

A generic estimator for the long-term error $\epsilon_2(i,t)$ may be defined by: 
\begin{equation}
\hat \mu_2(i,t) = \left( \frac{ Y_i(t)}{M_t} \right)^\star ~ \ \text{when} \ M_t>0,
\label{E:e2}
\end{equation}
where the $\star$ denotes the smoothing process, $Y_i(t)$ are the raw observations and $M_t=\underset{1 \leq i\leq N}{\text{med}} Z_i(t)$ is the median of the network. The $T$ transform from Equation~(\ref{E:mu-s}) is not required here, as we apply a moving average (MA) of length $L$ defined below.

This length $L$ should be larger than what is considered as short-term, that is, periods inferior to one solar rotation (27 days). Long-term on the other hand is usually defined as periods above 81 days~\citep{Dudok_gap}, beyond which the effects of the solar rotation and of the sunspot's lifetime are negligible. The mid-term temporal regime corresonds to periods between  27 and 81 days. To select the long-term scale for a given station $i$, we make the assumption that, for all $t$ belonging to a window of length $L$, we have:
\begin{equation}
\EE(\epsilon_2(i,t))=\mu_2(i,t) ~\simeq C_i,
\label{E:mu2C}
\end{equation}
where $C_i$ is a constant, which might differ from 1. Having $C_i=1$ means that the  station $i$ is at the same level as the median of the network. 
We test different lengths $L$ ($L>27$ days) for the MA window, and select the long-term regime as the shortest length for which the above assumption is valid. We consider thus $\hat \mu_2(i,t)$s of Equation~(\ref{E:e2}) in sliding windows  of length $L$ over the total period (1947-2013). 
We apply a non-parametric equivalent of the $t$-test (the Wilcoxon rank sum test~\citep{Wilcoxon}) on the $\hat \mu_2(i,t)$s to test whether Equation~(\ref{E:mu2C}) is verified within each window. Longer windows correspond thus to a stronger smoothing but also contain more values to test. As a result of this procedure, we define the long-term regime as all scales above 81 days as we found this to be the shortest length such that the constant assumption on $\mu_2(i,t)$ is not violated more than roughly $10\%$ of the time. This ties in with what solar physicists consider as the long-term regime.

Depending on our interest in detecting long-term drifts or jumps, different window lengths may be chosen in Equation~(\ref{E:e2}) (some well above 81 days). Indeed, drifts require long smoothing periods (several months, or even years) to be observed whereas  jumps might be oversmoothed by such long smoothing, and hence need a smaller MA window.

\begin{figure}[!htb]
  \centering
  \centerline{\includegraphics[scale=0.6]{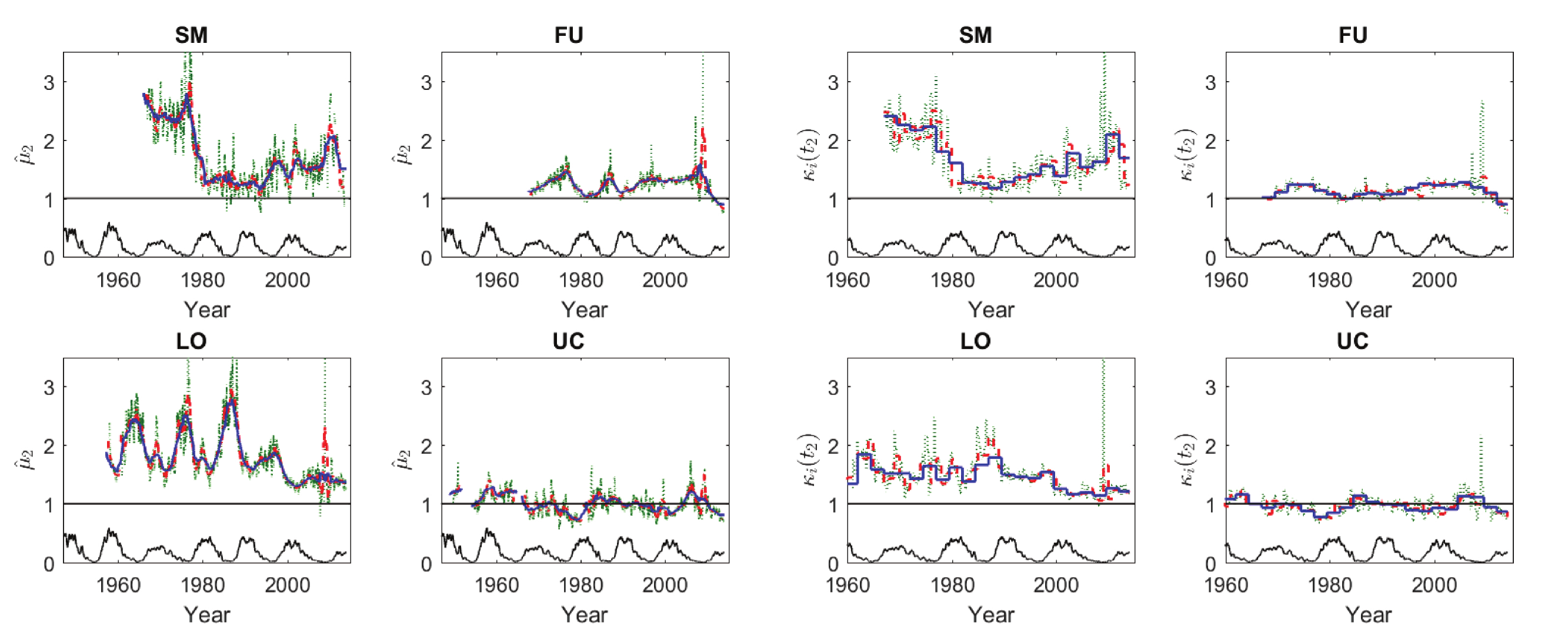}}
\caption{\small{(a)-(d): Estimation of $\hat \mu_2(i,t)$ for $N_s$ in four stations (SM, FU, LO and UC). $\hat\mu_2(i,t)$ is computed with different MA window lengths: 81 days (green dotted line), 1 year (red dashed line) and 2.5 years (blue plain line). 
(e)-(h): Estimation of the scaling-factors for $N_s$ in the same stations. The $\kappa_i(t_2)$s, with $1 \leq t_2 \leq T/\tau$, are computed using the $OLS(\vec{Y}_{i,t_2}|\vec{X}_{,t_2})$ regression in Equation~(\ref{E:ki}) on a block of $\tau=$ 81 days (green dotted line), 1 year (red dashed line) and 2.5 years (blue plain line). The solar cycle is represented in black at the bottom of the figures for $N_s$.}}
\label{fig:drifts}
\end{figure}

Figure~\ref{fig:drifts}(a)-(d) represent the long-term drifts associated to $N_s$ in four stations starting from 1960.  Figure~\ref{fig:drifts}(e)-(h) show the scaling-factors $\kappa_i(t_2)$s for the same stations used at short-term and minima regimes. We do not represent years before 1960 because FU and SM show too few observations in that period.  FU and UC appear relatively stable, unlike stations LO and SM, which display severe drifts. Bias in the counting process is also larger during solar minima when there are short-duration sunspots. This effect is clearly visible in LO. Indeed, erroneous encoding of counts leads to much higher relative errors during minima than during the remaining part of the solar cycle. Some jumps are also visible on the graphs with the smallest MA length (81 days) in green. This scale is more appropriate to observe the jumps, while longer scales only highlight the long-term drifts of the stations. \\ 

We emphasize here the  strong link between the pre-processing and the long-term analysis. 
Indeed, the scaling-factors presented in Section~\ref{sec:pre-processing} are a rough estimate of the long-term error, inspired by the historical procedure of J.R. Wolf. This rough estimation is required to rescale the stations to the same level. This rescaling is used to compute the median $M_t$ of Equation~(\ref{E:e2}). 
Contrarily to the piecewise constant $\kappa_i$s computed in Section~\ref{sec:pre-processing}, the $\hat \mu_2(i,t)$s are smooth over time and hence are more adapted to a future monitoring of the stations.

\subsection{Comparing stations with respect to their stability}
\label{S:error_comparing}

In previous sections, we presented separately the   estimations of the short-term error $\hat{ \widetilde{\epsilon}}(i,t)$, the long-term error $\hat \mu_2(i,t)$ and the error at minima $\hat \epsilon_3(i,t)$. All three types of errors  are needed to assess the quality and stability of one station. 
It is more important for a station to have a low variability (low interquantile range) rather than to be aligned on the mean on the network. Indeed, as seen in Section~\ref{sec:pre-processing}, it is easy to rescale a station on the mean of the network.

\begin{figure}[!htb]
  \centering
  \centerline{\includegraphics[width=9cm, height=7cm]{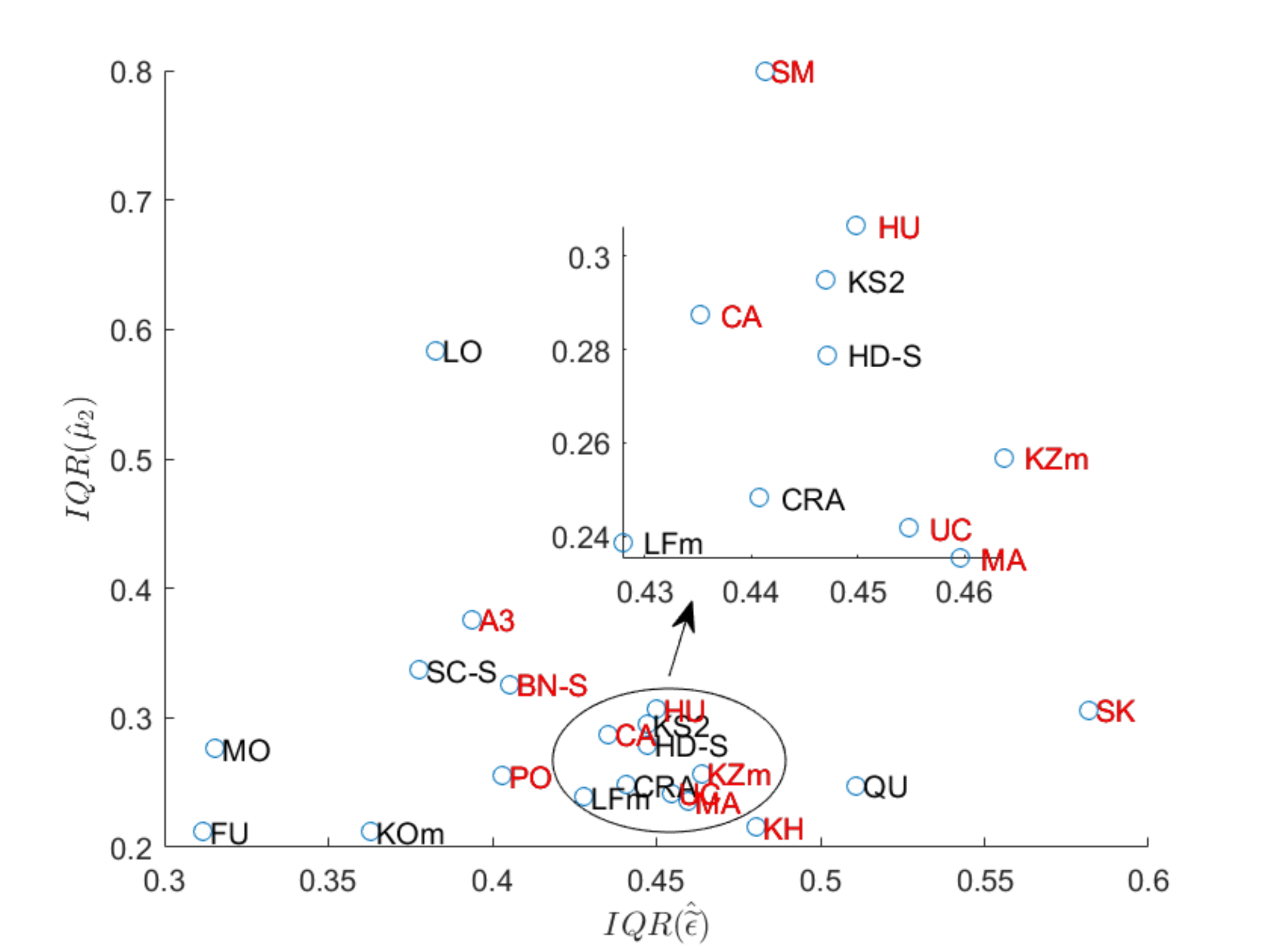}}
\caption{\small{Scatter plot showing the interquantile range of the estimated short-term error $\hat{ \widetilde{\epsilon}}(i,t)$ and the interquantile range of the estimated long-term error $\hat \mu_2(i,t)$, station by station. Stations in red represent the teams of observers, the others are single observers.}}
\label{fig:tutti}
\end{figure}

Figure~\ref{fig:tutti} displays a visual representation of long-term against short term error  for each station. It shows the long-term versus short-term empirical interquantile range on a 2D plot, and characterizes thus the stability of the stations outside of minima. 
Stations in red are the teams of observers. They usually experience more short-term variability than an individual.
We see that MO (Mochizuki, Japan), FU and KOm (Koyama, Japan) have low-variability both on short and long-term. They correspond to long individual observers with stable observation practises. On the other hand, the LO station shows a poor long-term stability, while its short-term variability is remarkably low for a professional observatory. As mentioned earlier, this is due to the fact that there is a main observer. UC shows a large variability in the short-term (due to many observers) but an interesting long-term stability, as already noticed in Figure~\ref{fig:e1-station}. SM experiences the most severe long-term variability of the network. It has also a large short-term variability, characteristic of a team of observers.  QU shows a large short-term variability and a low long-term variability level. Although it seems it is a single observer, it appears there was a move from one place to another during the observing period, and maybe a change of instrument that would impact the short term variability.  This surprising behaviour will prompt SILSO to ask for more metadata. 

\section{CONCLUSION AND FUTURE PROSPECTS}
\label{sec:conclusion}

In this article, we propose the first comprehensive uncertainty model in a \emph{multiplicative} framework for counting spots, groups and composite on the Sun. Our approach is robust to missing values and was applied on 66 years of data (1947-2013).  We presented several parametric models for the density of the \lq true'  $N_s$, $N_g$ and $N_c$, as well as for the density of their error distribution at minima, short, and long-term.  This error quantitication allows proposing a first classification of the 21 stations of our pool based on their stability.  It shows that the observatories are affected differently by the various type of errors: some are stable with respect to the network at short-term but experience large drifts, and conversely.  The analysis highlights the hazards of using a single-pilot station as the unique reference of the network. 

We intend to use the error models presented in Section~\ref{sec:errors} for a parametric monitoring of all stations of the network, with a particular focus on new stations. Data from new-born observatories can be recorded for several months. Their distributions may then be compared to the density of the short-term error (or the error at minima if we are in a minima period) of the entire network obtained in this paper. If the stations experience similar errors, they may be included in the network. Otherwise, the stations might need to improve or correct their observing procedure before entering the SILSO network. 

A non-parametric monitoring that aims to detect in quasi real-time the long-term drifts of the stations in the network is also under development.  An example of a classical monitoring procedure is the CUSUM chart~\citep{CUSUM}. It is frequently used to control the production quality in industry. The chart accumulates the deviations of the mean value above a reference level in a statistic. If the value of the statistic exceeds a pre-defined threshold depending on the standard deviation of the process, the process is considered out-of-control and an alert is given. This simple method based on the two first moments of the distribution is obviously not adequate to control heavily-skewed variables such as the number of spots. More complex methods need to be developed that will strongly depend on the models of the data.

The work presented here enhances our comprehension of the ISN and its error.
It is part of a larger project that aims at improving the quality of the ISN. We started this project a few years ago when a revised version of the ISN was published~\citep{all_corrections}, and we will pursue with the future monitoring of the stations to provide a yet missing quality-control procedure for the ISN. As the ISN is used as a benchmark in several fields of astrophysics and space physics, this is a much needed task.

\section{ACKNOWLEDGMENT}
\label{sec:acknowledgment}
This work benefited from highlighting discussions with T. Dudok de Wit. 
The first author gratefully acknowledges funding from the Belgian Federal Science Policy Office (BELSPO) through the BRAIN VAL-U-SUN project (BR/165/A3/VAL-U-SUN).

\bibliography{sunspot_complete}


\newpage 
\appendix

\section{Time-scales of the pre-processing}
\label{sec:AA}

This appendix details the statistical procedure selecting the time-scales of the pre-processing described in Section~\ref{sec:pre-processing}.
It is composed of three steps. First, the daily scaling-factors are computing using: 
 \begin{equation}
 \kappa_i((t_1, t_2))=\frac{Y_i((t_1, t_2))}{\text{med}_{1 \leq i\leq N} Y_i((t_1, t_2))} 
\label{E:ki_sum}
\end{equation}
where, as in Section~\ref{sec:pre-processing}, we rewrite the time by an array of two indices $1 \leq t_1 \leq 30$ and $1 \leq  t_2 \leq T/30$, corresponding respectively to the day and the month of the observation.

\begin{figure}[htbp]
	\centering
		\includegraphics[width=11cm, height=8cm]{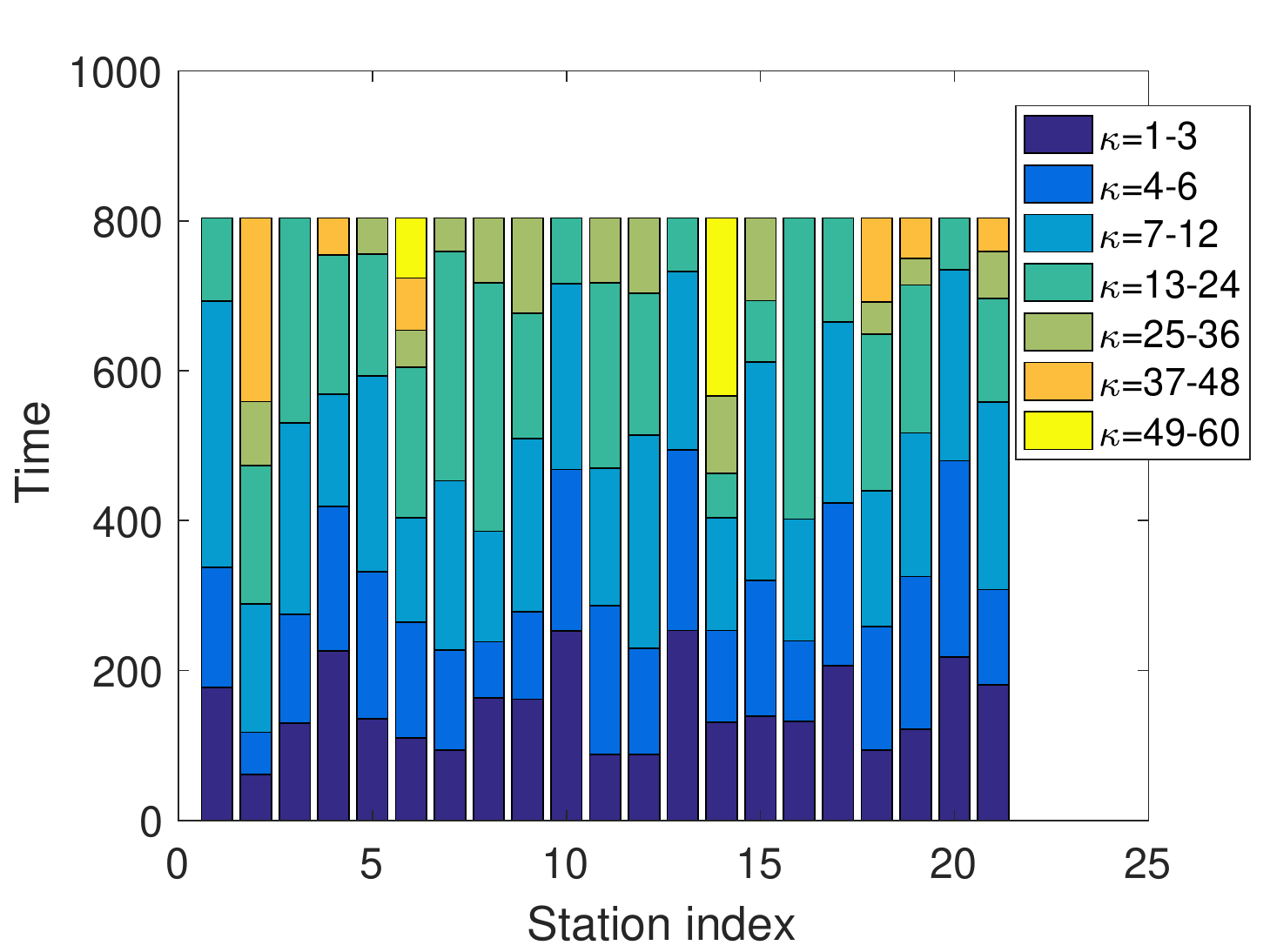}
	\caption{\small{Bar-chart representing the results of the KW test applied to the scaling-factors for $N_s$.
	The x-axis represents the stations indexed from 1 to 21 and the y-axis shows the total period studied expressed in months (1 unit $\approx$ 30 days). The y-axis is not ordered in time, for readability purpose, but it is ordered with respect to the length of the segments. The colours of the chart correspond to the number of blocks that may be grouped into a single factor (\lq $\kappa=5$' means that a single scaling-factor may be computed for a period of 5 months). }}
\label{fig:segmentation}
\end{figure}

 Second, the non-parametric Kruskal-Wallis test (KW)~\citep{Kruskal_Wallis} is applied on blocks of 30 factors, since the \lq k-coefficients' of Equation~(\ref{E:k_factor}) are currently estimated on a monthly basis at the WDC-SILSO. Let $\vec{\kappa}_{i,t_2}=[\kappa_i((t_1,t_2))]_{1 \leq t_1 \leq 30}$ denote the vector of the daily factors on one month.
The test assesses whether the $\vec{\kappa}_{i,t_2}$s of consecutive months are statistically different. The procedure starts by comparing the distribution of the first month of the period studied, $\vec{\kappa}_{i,1}$,  to the distribution of the second month, $\vec{\kappa}_{i,2}$.
 If the test shows that both distributions are significantly different, the two next distributions $\vec{\kappa}_{i,2}$ and $\vec{\kappa}_{i,3}$ are tested. Otherwise, the distribution of the two first months [$\vec{\kappa}_{i,1}$ $\vec{\kappa}_{i,2}$] is compared to the distribution of the third month $\vec{\kappa}_{i,3}$. 
The algorithm is iterated until the end of the period, for each station. 
Note that the KW test performs well when comparing two or more independent samples of unequal sizes.  
The correlation of the data is thus neglected in this procedure. Despite the presence of correlations between consecutive days, the correlation between consecutive months is low.   
The test provides thus a station-specific segmentation, shown in Figure~\ref{fig:segmentation} for $N_s$.  The length of the segments indicates the number of consecutive blocks of scaling-factors that come from the same distribution. We assume that these factors are constant within each segment.  

In the last step, the global time-scales for each index are defined from the segmentations of the individual stations.  
The length of the most frequent segment is first selected in each station. Then a global scale is estimated from the median of the most frequent lengths by station, for $N_s$, $N_g$ and $N_c$.

\end{document}